\documentclass[12pt]{article}
\usepackage[margin=1in, paperwidth=8.5in, paperheight=11in]{geometry}
\usepackage{amsmath}
\usepackage{graphicx}%
\usepackage{amsfonts}%
\usepackage{amssymb}
\usepackage{color}
\usepackage{float,subfigure}
\usepackage{setspace}
\usepackage{comment}
\usepackage{setspace}     
\usepackage{chngpage}  
\usepackage{epsfig}
\usepackage{subfigure}
\usepackage{hyperref}
\usepackage{multirow}
\usepackage{ifsym}
\usepackage{caption}
\usepackage{wrapfig}
\usepackage{appendix}
\usepackage{layout}
\usepackage{titlesec}
\usepackage{rotating}
\usepackage{array,arydshln} 
\usepackage{amssymb}
\usepackage{pifont}
\usepackage[flushleft]{threeparttable}
\usepackage{enumitem}
\usepackage{flafter}

\titlespacing\section{0pt}{12pt plus 4pt minus 2pt}{0pt plus 2pt minus 2pt}
\titlespacing\subsection{0pt}{12pt plus 4pt minus 2pt}{0pt plus 2pt minus 2pt}

\date{}
\newcommand{\bet}{ \mbox{\boldmath $ \eta $} }

\newcommand{\bbeta}{ \mbox{\boldmath $\beta$} }
\newcommand{\bb}{ {\bf b} }

\newcommand{\bx}{ {\bf x} }
\newcommand{\bX}{ {\bf X} }

\newcommand{\bR}{ {\bf R} }

\newcommand{\br}{ {\bf r} }

\newcommand{\ind}{\perp\!\!\!\!\perp}

\usepackage[natbib=true,
    style=authoryear,
    sorting=none]{biblatex}
\numberwithin{equation}{section}

\begin{document}

\thispagestyle{empty}
\setcounter{page}{0}

\bigskip
\begin{center}
{\large \textbf{Network Meta-Analysis of Time-to-Event Endpoints with Individual Participant Data using Restricted Mean Survival Time Regression}}

\bigskip
\bigskip

\textbf{Kaiyuan Hua$^{1}$, Xiaofei Wang$^{1}$, Hwanhee Hong$^{1}$}\\
$^{1}$ Department of Biostatistics and Bioinformatics, Duke University School of Medicine, Durham, North Carolina, 27705, USA \\

\singlespacing
\end{center}

\small
\doublespacing
\textsc{Abstract:} 
Restricted mean survival time (RMST) models have gained popularity when analyzing time-to-event outcomes because RMST models offer more straightforward interpretations of treatment effects with fewer assumptions than hazard ratios commonly estimated from Cox models. However, few network meta-analysis (NMA) methods have been developed using RMST. In this paper, we propose advanced RMST NMA models when individual participant data are available. Our models allow us to study treatment effect moderation and provide comprehensive understanding about comparative effectiveness of treatments and subgroup effects. An extensive simulation study and a real data example about treatments for patients with atrial fibrillation are presented.

\bigskip

\textsc{Key words: Effect moderation; Individual participant data; Network-meta analysis; Restricted mean survival time.}

\section{Introduction} \label{s:intro}
Network meta-analysis (NMA) is an extension of traditional pairwise meta-analysis~\citep{dersimonian1986meta} and evaluates more than two treatments simultaneously by integrating findings from multiple trials, particularly randomized controlled trials (RCTs), and accounting for heterogeneity across the trials~\citep{lu2004combination,salanti2008evaluation,mills2013demystifying,white2015network,hong2016bayesian}. With limited evidence about head-to-head comparisons in RCTs, NMA can leverage information about both ``direct" and ``indirect" comparisons of treatment, providing comprehensive and integrative inferences about the comparative effectiveness of treatments~\citep{Saramago14,lu2004combination}. NMA often combines aggregate data (e.g., summary statistics), denoted by NMA-AD, but thanks to recent advances in data sharing and increasing access to individual participant data (IPD), statistical methods for NMA using IPD, denoted by IPD-NMA, have been actively developed~\citep{donegan2013combining,Hong15,debray2015get}. Using IPD in NMA allows researchers to inform personalized treatment decisions by assessing treatment effect moderation and subgroup effects using individual-level covariates~\citep{lyman2005strengths,riley2010meta,donegan2012assessing,Hong15}.

There are two approaches for conducting IPD-NMA: two-stage and one-stage approaches~\citep{simmonds2005meta}. The two-stage approach first estimates treatment effects within each study, and then combines the study-level effect estimates across studies using a NMA-AD model~\citep{debray2013individual}. On the other hand, the one-stage approach combines the IPD of all studies included in the NMA simultaneously as a single step by using a hierarchical model~\citep{bowden2011individual}. The differences between these two methods have been extensively discussed through empirical, theoretical, or simulation studies by~\cite{stewart2012statistical,debray2013individual,burke2017meta,hua2017one}, and~\cite{kontopantelis2018comparison}. The benefits of using the one-stage methods include: 1) they use an exact likelihood specification instead of assuming the asymptotic normality of study estimates that are usually used in the two-stage methods; 2) they can account for within-study parameter correlation more efficiently, especially in the presence of missing outcomes; 3) the one-stage methods provide higher power and lower bias when estimating treatment-by-covariate interactions to assess effect moderation; 4) the one-stage methods can provide reliable results with limited numbers of studies and study participants or events per study are available. However, the one-stage methods may entail a greater computational burden compared to the two-stage methods, and the two-stage methods can be used when a few studies provide IPD and the remaining studies provide aggregate data.

When analyzing a time-to-event outcome in NMA, using IPD has many advantages~\citep{Smith05}. First, we can apply the same outcome model and assess model assumptions consistently across trials. Second, a common metric quantifying treatment effects with respect to time-to-event outcomes (e.g., hazard ratio or restricted mean survival time) can be estimated directly in all available trials in IPD-NMA regardless of what was reported in publications. Third, the individual-level covariates enable us to assess treatment effect moderation accurately compared to aggregate covariates. Many statistical methods for time-to-event outcomes typically rely on hazard ratios (HRs), which is usually estimated by Cox proportional hazards (PH) models~\citep{cox1972regression}. NMA with time-to-event outcomes is not an exception and IPD-NMA models based on Cox regression have been proposed by~\cite{Smith05} and~\cite{crowther2012individual}.
However, Cox regression models rely heavily on the PH assumption and could result in misleading HR estimates and inappropriate summary of treatment effect when the assumption is violated~\citep{lin1989robust}.

As an alternative to HR, restricted mean survival time (RMST) was proposed by~\cite{irwin1949standard} and popularized by many researchers~\citep{andersen2004regression,royston2013restricted,uno2014moving,tian2014predicting}. Let $T$ be the event time and $t^*$ be a clinically relevant pre-specified truncation time at which we want to evaluate the treatment effect. The RMST, $\mu(t^*)$, is defined as the mean of the truncated event time $Y = T \wedge t^*$, where $\wedge$ represents the minimum of two values. The RMST is calculated by the area under the survival curve $S(t)$ from $t = 0$ to $t = t^*$ written as follows~\citep{royston2013restricted}:
\[
\mu(t^*) = E(Y) = E[T \wedge t^*] = \int_0^{t^*}S(t)dt.
\] 
Typically a nonparametric estimation of $\mu(t^*)$ can be obtained by 
$\int_0^{t^*}\hat{S}(t)dt$, where $\hat{S}(t)$ is the Kaplan-Meier estimator of the survival function. There are also several other methods to calculate RMST such as parametric survival models~\citep{royston2002flexible,wei2015meta}, pseudo-value regression method~\citep{andersen2004regression}, and inverse probability of censoring weighted regression method~\citep{tian2014predicting}. Based on RMST, a relative treatment effect between two groups (e.g., active treatment versus control) can be estimated by a difference in RMSTs between the two groups, which captures the gain or loss of mean event-free survival time up to $t^*$~\citep{kim2017restricted}. 

RMST has more straightforward clinical interpretation than HRs and provides a valid measure for treatment effects under any distributions of time-to-event outcomes with no model assumptions~\citep{perego2020utility}. Another advantage of RMST is its ability to estimate treatment effects at different time points, avoiding the conceptual dilemma of averaging HRs over time especially when survival curves between two groups cross each other, an indication of PH assumption violation.

Recently, RMST methods have been incorporated into NMA with IPD~\citep{wei2015meta,lueza2016difference,weir2021multivariate,tang2022bayesian}. A two-stage IPD-NMA model is commonly used where the first stage obtains RMST estimates per treatment arm in each study and then the second stage conducts a traditional study-level NMA using the RMST estimates. However, this model does not adjust for covariates in the first stage properly, because RMSTs are estimated by a simple integration of a survival curve, $\int_0^{t^*}\hat{S}(t)dt$, nonparametrically in the first stage. Moreover, the existing methods use a contrast-based parameterization that directly models relative treatment effects such as an RMST difference between two treatments. As an alternative, an arm-based parameterization can directly model the responses observed in each of the treatment arms~\citep{hawkins2016arm}, and it enables direct modeling of RMSTs rather than RMST differences in NMA, which provides straightforward parameter interpretations~\citep{white2019comparison}. By estimating RMSTs, we can calculate the RMST differences for all pairwise comparisons of treatments.

To the best of our knowledge, there is no formal model development for one-stage IPD-NMA using RMST. Moreover, if covariates are properly used to estimate treatment effect in the first stage, two-stage model would still be able to adjust covariate when estimating treatment effect. As such, in this paper, we propose advanced two-stage and one-stage IPD-NMA methods for time-to-event outcomes using the arm-based parameterization to estimate RMST. Our methods fully utilize individual-level covariates to adjust RMST by covariates and estimate treatment-by-covariate interaction effects.

The remainder of this paper is structured as follows. The Section~\ref{s:methods} introduces our proposed RMST IPD-NMA methods. In Section~\ref{s:simulation}, results from an extensive simulation study is provided. In Section~\ref{s:data analysis}, we illustrate the proposed methods using a real NMA example about treatments for patients with atrial fibrillation~\citep{carnicelli2021individual,carnicelli2022direct}. We conclude the paper with discussions and limitations of the proposed methods in Section~\ref{s:discussion}.

\section{Methods} \label{s:methods}
We propose two RMST IPD-NMA models under (1) two-stage and (2) one-stage approaches. Both models estimate RMST adjusted by covariates using the inverse probability of censoring weighted regression method~\citep{tian2014predicting}. The models are developed under the consistency assumption, where direct and indirect comparisons of treatments are the same~\citep{higgins2012consistency}.

\subsection{Inverse Probability of Censoring Weighted RMST Regression} \label{s:methods.1}
We first provide a brief introduction of the inverse probability of censoring weighted RMST regression model that is used to estimate RMSTs adjusted by covariates. Suppose that $T$ is a time-to-event and $\Tilde{\bX}$ is a $p \times 1$ covariate vector for a study participant, where $\Tilde{\bX}$ can contain any factors including intercept, treatment indicators, baseline covariates, and treatment-by-covariate interactions. Here, $T$ is subject to right censoring with censoring time $C$, and we assume conditionally independent censoring, formally written as $C \ind T | \Tilde{\bX}$. Let $U = T \wedge C$ and $\delta = I[T \leq C]$ be the censoring indicator. For a specific truncation time $t^*$, we let $Y = T \wedge t^{*}$ be the truncated individual time-to-event outcome at $t^*$. Then, the RMST up to $t^*$ given $\Tilde{\bX} = \Tilde{\bx}$ is defined by $\mu(t^{*}|\Tilde{\bx}) = E(T \wedge t^{*}|\Tilde{\bx}) = E(Y|\Tilde{\bx})$. In a study with $n$ participants, the observed data are written as $\{(U_i, \delta_i, \Tilde{\bX}_i), i = 1, \dots, n\}$. Although $Y_i$ may not be observable for all participants due of censoring, its expected value is still estimable~\citep{tian2014predicting}.

The RMST regression model is written as $g(\mu(t^{*}|\Tilde{\bx})) = \Tilde{\bx}^{T}\bbeta$, where $\bbeta$ is the vector of regression coefficients and $g(\cdot)$ is the link function (e.g., log link or identity link). Since the distribution of $Y$ is unknown, a quasi-likelihood method is applied to estimate model parameters~\citep{tian2014predicting,gardiner2021restricted}. We estimate $\hat{\bbeta}$ by solving the following estimating equation:
\[
S_n(\hat{\bbeta}) = n^{-1}\sum_{i=1}^n w_i \Tilde{\bx}_i \left\{ Y_i - g^{-1}(\Tilde{\bx}_i^{T}\hat{\bbeta}) \right\} = 0.
\]
Here, $w_i = \delta_i^*/\hat{G}(Y_i)$ is the inverse probability censoring weight for the $i^{th}$ participant, where $\delta_i^* = I[C_i \geq Y_i]$ and $\hat{G}(t) = Pr(C > t)$, the Kaplan-Meier estimator of the censoring time $C$.

\subsection{Two-Stage RMST IPD-NMA Model} \label{s:methods.2}
The first stage of the two-stage RMST IPD-NMA model analyzes individual study separately to estimate RMSTs and treatment-by-covariate interaction effects by fitting an inverse probability of censoring weighted RMST regression as introduced in Section~\ref{s:methods.1}. Suppose the time-to-event outcome is $T_{ij}$ and the censoring time is $C_{ij}$, where $i$ and $j$ index participants and studies such as $i=1,\dots,n_j$ and $j = 1,\dots,J$. The truncated time-to-event outcome is $Y_{ij} = T_{ij} \wedge t^*$. Suppose we evaluate $K$ treatments and $Z_{ij} \in \{1,\dots,K\}$ is the treatment indicator. We consider $P$ covariates, where the $p^{th}$ covariate is $X_p$, which can be either discrete or continuous, and we let $\bX_{ij}$ be a $P\times 1$ covariate vector specific for Patient $i$ in Study $j$. Let $\mu_{ij} = E(Y_{ij} | \bX_{ij}, Z_{ij})$ be the conditional RMST, then the RMST regression model for the $j^{th}$ study in the first stage is written as follows:
\begin{eqnarray}  \label{eqn:1}
    g(\mu_{ij}) = \sum_{k=1}^K\alpha_{jk}I[Z_{ij}=k] + \sum_{k=1}^K\bX_{ij}\bbeta_{jk}I[Z_{ij}=k].
\end{eqnarray}
Typical choices of the link function $g(\cdot)$ can be log link or identity link. We will use the log link function (i.e., $g(\cdot) = log(\cdot)$) throughout the paper. Note that our model adopting the arm-based parameterization does not need a separate intercept or main covariate effect parameters in the model~\citep{Hong15,hong2016bayesian}. Here, $\alpha_{jk}$ is the log-RMST for Treatment $k$ and $\bX_{ij}=0$ in the $j^{th}$ study. We define $\bbeta_{jk} = (\beta_{jk1},\dots,\beta_{jkP})^T$, where $\beta_{jkp}$ is the study-specific treatment-by-covariate interaction effect for $p = 1,\dots,P$. The resulting estimates in the $j^{th}$ study are denoted as $(\hat{\alpha}_{j1},\dots,\hat{\alpha}_{jK},\hat{\bbeta}_{j1}^{T},\dots,\hat{\bbeta}_{jK}^{T})$ and their associated study-specific variance-covariance matrix $\hat{\Sigma}_j$, where $\hat{\bbeta}_{jk} = (\hat{\beta}_{jk1},\dots,\hat{\beta}_{jkP})^T$.

In the second stage, the estimated log-RMSTs, $(\hat{\alpha}_{j1},\dots,\hat{\alpha}_{jK})$, and treatment-by-covariate interactions, $(\hat{\bbeta}_{j1}^{T},\dots,\hat{\bbeta}_{jK}^{T})$, are analyzed using a multivariate meta-analysis model~\citep{chen2012method,schwarzer2015meta}. The multivariate meta-analysis model is written as follows: 
\begin{equation} \label{eqn:2}
\begin{aligned}
    \hat{\alpha}_{jk} & = \alpha_k + a_{jk} + \epsilon_{jk}, \\
    \hat{\beta}_{jkp} & = \beta_{kp} + b_{jkp} + \eta_{jkp},
\end{aligned}
\end{equation}
for $j=1,\dots,J$, $k = 1,\dots,K$, and $p = 1,\dots,P$. Here, $\alpha_k$ is the fixed (mean) main treatment effect, interpreted by the log-RMST up to $t^*$ of Treatment $k$ for $\bX_{ij}=0$, and $\beta_{kp}$ is the fixed (mean) treatment-by-covariate interaction effect between Treatment $k$ and $X_p$. If $X_p$ is a binary covariate, $\beta_{kp}$ is interpreted as the difference in log-RMST for Treatment $k$ between $X_p=1$ and $X_p=0$. If $X_p$ is a continuous covariate, $\beta_{kp}$ is interpreted as the change in log-RMST for Treatment $k$ per unit change in $X_p$. Both $a_{jk}$ and $b_{jkp}$ are associated with random effects $\alpha_k$ and $\beta_{kp}$, respectively. Let $\bb_{jk} = (b_{jk1},\dots,b_{jkP})^T$. We assume that the random effects parameter vector, $\br_j = (a_{j1},\dots,a_{jK},\bb_{j1}^T,\dots,\bb_{jK}^T)$, independently follows a multivariate normal distribution, namely $MVN(0, \bR_2)$. Here $\bR_2$ is the unstructured covariance-variance matrix of the random effects that describes the between-study heterogeneity for each parameter in Equations~\eqref{eqn:2} and the diagonal element of $\bR_2$ is the variance component of the between-study heterogeneity corresponding to each parameter. In addition, $\epsilon_{jk}$ and $\eta_{jkp}$ are the associated residuals for $\alpha_k$ and $\beta_{kp}$, respectively, which are assumed to be independent. Let $\bet_{jk} = (\eta_{jk1},\dots,\eta_{jkP})$, then the residual vector $(\epsilon_{j1},\dots,\epsilon_{jK},\bet_{j1}^T,\dots,\bet_{jK}^T)$ is assumed to follow a multivariate normal distribution with mean 0 and covariance-variance matrix $\hat{\Sigma}_j$, i.e. $MVN(0, \hat{\Sigma}_j)$.

We use the restricted maximum likelihood method~\citep{jackson2011multivariate} to fit the models in Equations~\eqref{eqn:2}, implemented using R package \texttt{mvmeta} (version 1.0.3)~\citep{gasparrini2015package}. Alternatively, one can also use a multivariate version of the DerSimonian-Laird method of moments estimator~\citep{jackson2013matrix} or Bayesian approach~\citep{wei2013bayesian} to fit this multivariate model with random effects.

\subsection{One-Stage RMST IPD-NMA Model} \label{s:methods.3}
In the one-stage model, IPD from all studies included in NMA are modeled simultaneously. The inverse probability of censoring weighted RMST regression model is extended to combine multiple studies and account for between-study heterogeneity. We use the penalized quasi-likelihood (PQL) method~\citep{breslow1993approximate} to fit the one-stage model with random effects as our individual truncated event time outcome does not follow a normal distribution. Followed by the same notations as in the two-stage RMST IPD-NMA model, given the study level random effects, $Y_{ij}$ are assumed to be conditionally independent with the mean $E(Y_{ij}|\br_j) = \mu_{ij}^{(\br_j)}$ and the variance $\mbox{Var}(Y_{ij}|\br_j) = \phi v(\mu_{ij}^{(\br_j)})/w_{ij}$. Here, $\mu_{ij}^{(\br_j)}$ is related to the value of random effects $\br_j = (a_{j1},\dots,a_{jK},\bb_{j1}^T,\dots,\bb_{jK}^T)$, $\phi$ is the dispersion parameter and $w_{ij} = \delta_{ij}^*/G(Y_{ij})$ is the inverse probability of censoring weight. In addition, by assuming the link function $g(\cdot)$ is a canonical link~\citep{mccullagh2019generalized}, $v(\cdot) = 1/g'(\cdot)$, so that $v(\mu) = \mu$ for log link. The one-stage RMST IPD-NMA model is written as follows:
\begin{equation} \label{eqn:3}
      g(\mu_{ij}^{(\br_j)}) = \sum_{k=1}^K\alpha_{jk}I[Z_{ij}=k] + \sum_{k=1}^K\bX_{ij}\bbeta_{jk}I[Z_{ij}=k].
\end{equation}

For $k=1,\dots,K$, we define $\alpha_{jk} = \alpha_k + a_{jk}$ and $\bbeta_{jk} = (\beta_{jk1},\dotso,\beta_{jkP})^T$, where $\beta_{jkp} = \beta_{kp} + b_{jkp}$ ($p = 1,\dotso,P$). Here, $\alpha_k$, $a_{jk}$, $\beta_{kp}$, and $b_{jkp}$ have the same interpretations as in Equation~\eqref{eqn:2}. Let $\bb_{jk} = (b_{jk1},\dots,b_{jkP})$, we assume the random effects parameter vector, $\br_j$, independently following a multivariate normal distribution with mean 0 and an unstructured covariance-variance matrix $\bR_1$, namely $MVN(0, \bR_1)$. Here, $\bR_1$ is the covariance matrix of the random effects in Equation~\eqref{eqn:3} and each diagonal element of $\bR_1$ measures the variance of the between-study heterogeneity. The PQL method is implemented using the \texttt{glmmPQL} function available under R package \texttt{MASS} (version 7.3-58.3)~\citep{ripley2013package}. We show the algorithm for calculating the parameters in Equation~\eqref{eqn:3} in Web Appendix A in the Supplementary Material.

In both two-stage and one-stage models, including too many covariates can result in numerical difficulties when estimating random effects for all coefficients. In such cases, there are several ways to simplify the models. First, if we expect no interaction effect between covariate $X_p$ and the treatments, we can replace $\beta_{jkp}$ with $\beta_{jp}$ in Equation~\eqref{eqn:1} and~\eqref{eqn:3}. Second, random effects can be removed for some parameters if there exists little or no between-study variability. Third, covariance matrices for random effects, $\bR_2$ and $\bR_1$, can be assumed to be structured such as diagonal or blocked diagonal, rather than unstructured.

\subsection{Existing Methods} \label{s:methods.4}

We consider two existing two-stage IPD-NMA proposed by~\cite{weir2021multivariate} using a frequentist framework (denoted by nonparametric frequentist [NPF] two-stage model) and~\cite{tang2022bayesian} using a Bayesian framework (denoted by nonparametric Bayesian [NPB] two-stage model). We modify these models by employing the arm-based parameterization so that their results can be compared with those under our methods. The first stage of the two models estimates the study-specific RMSTs for a treatment using $\int_0^{t^*}\hat{S}(t)dt$ and the associated variance using $\int_0^{t^*}\{\int_t^{t^*}\hat{S}(u)du\}^2\frac{d\hat{\Lambda}(t)}{Y(t)}$. Here, $\hat{S}(t)$ is the Kaplan-Meier estimator of the survival function, $\hat{\Lambda}(t) = -log\{\hat{S}(t)\}$ is the Nelson-Aalen estimator, and $Y(t)$ is the number of patients at risk at time $t$. For the NPF model, the second stage combines the study-specific RMSTs by using a standard NMA-AD approach. The modified NPF model under the arm-based parameterization is fitted using the multivariate meta-analysis model~\citep{chen2012method,schwarzer2015meta} yielding restricted maximum likelihood estimator. For the modified NPB model, we adapt a Bayesian arm-based model proposed by~\cite{Hong15,hong2016bayesian} to combine the study-specific RMSTs. Note that the existing models are not designed to study effect moderators because the first stage of these models uses a simple integration of the marginal survival function.

\section{Simulation Study} \label{s:simulation}
In this simulation study, we evaluate finite sample performance of the proposed two-stage and one-stage RMST IPD-NMA models under various scenarios, and compare the proposed methods with the existing methods in terms of bias, mean squared error (MSE) and coverage probability of the RMST estimators. We use 1,000 Monte Carlo replications. The coverage probability is defined as the proportion of the number of replications of which 95\% confidence intervals (or credible intervals when fitting NPB model) contain the true value out of the total number of replications. Our data generating process is built based upon the approaches proposed by~\cite{wang2018modeling} and~\cite{zhong2022restricted}.

\subsection{Setting} \label{s:simulation1}
We simulate data for a network meta-analysis of randomized trials to assess three treatments: Treatment A ($k=1$), Treatment B ($k=2$) and Treatment C ($k=3$). We generate the time-to-event outcome $T_{ij}$ using the following model:
\begin{eqnarray} \label{eqn:4}
log(T_{ij}) = \sum_{k=1}^3\alpha_{jk}^*I[Z_{ij}=k] + X_{ij}\sum_{k=1}^3\beta_{k}^*I[Z_{ij}=k] + \epsilon_{ij}\sum_{k=1}^3\sigma_k,
\end{eqnarray}
where $X_{ij} \sim Bern(0.5)$, a binary covariate. Here, we sample $\alpha_{jk}^*$ from $N(\alpha_k^*,\tau^2)$, where $\tau$ measures heterogeneity across studies. We consider three values of $\tau = 0.1, 0.3, 0.5$ indicating low, moderate, and high between-study heterogeneity. Note that Equation~\eqref{eqn:4} is an accelerated failure time (AFT) model with random effects, where $\alpha_k^*$ is the fixed treatment effect for $X_{ij}=0$, $\beta_k^*$ is the fixed treatment-by-covariate interaction effect, $\sigma_k$ is a scale parameter, and we assume the residual $\epsilon_{ij} \sim N(0,1)$. The prespecified values of $\alpha_k^*$, $\beta_k^*$, and $\sigma_k$ are provided in Web Table 1 in the Supplementary Material.

Next, we assume the censoring time $C_{ij}$ is independent from $T_{ij}$, which is generated from an exponential distribution $Exp(0.15)$. We set a fixed truncation time for RMST at $t^*=4$. Web Figure 1 in the Supplementary Material displays the simulated survival curves in the two subgroups, $X=0$ and $X=1$, for three treatments.

We fit the proposed two-stage and one-stage RMST IPD-NMA models with the $log$ link function. The estimands are the log-RMSTs for each treatment and subgroup. The true log-RMSTs are calculated as 0.687, 1.070, 0.877 for Treatment A, B, and C in subgroup $X=0$ , and 0.859, 1.186, and 1.056 for Treatment A, B, and C in subgroup $X=1$. We have a discussion about the true estimand calculation in Web Appendix B in the Supplementary Material.

In this simulation, we vary one of the three factors while fixing the others. The three factors include: (1) the total sample size of NMA data; (2) the number of trials included in an NMA; and (3) the network of treatments. Let $nt$ and $n$ denote the number of trials and the sample size in each trial, respectively. The simulated NMA data structures are determined by the following setups: \\
Scenario 1: Vary $n = 200, 500, \mbox{or } 1000$ while fixing $nt=20$. All trials have three treatment arms. \\
Scenario 2: Vary $nt = 10, 20, \mbox{or } 30$ while fixing $n=500$. All trials have three treatment arms.\\
Scenario 3: Vary treatment networks (shown in Table~\ref{tab:1}) while fixing $nt=20$. Let the sample size in each trial be sampled from a discrete uniform distribution, $n \sim Unif(300,700)$.

Note that we assume an equal allocation of treatments in each trial. In Scenarios 1 and 2, we evaluate bias, MSE and coverage probability of RMSTs when varying  $nt$ and $n$, respectively, while fixing the other parameter. Scenario 3 considers realistic treatment networks, where no trials compare all three treatments simultaneously (i.e., no three-arm trials). In Scenario 3, the three networks result in different numbers of trials included in NMA. For example, 18 trials study Treatment A in Network 1, while 14 trials study Treatment A in Network 3 (see the bottom of Table 1). That is, Network 1 contains more data (i.e., information) about Treatment A than Network 3. Note that Scenario 3 is designed as the number of trials studying each treatment decreases from Network 1 to Network 3.

\subsection{Results} \label{s:simulation2}
\begin{figure}[!htbp]
\centering\includegraphics[scale = 0.6]{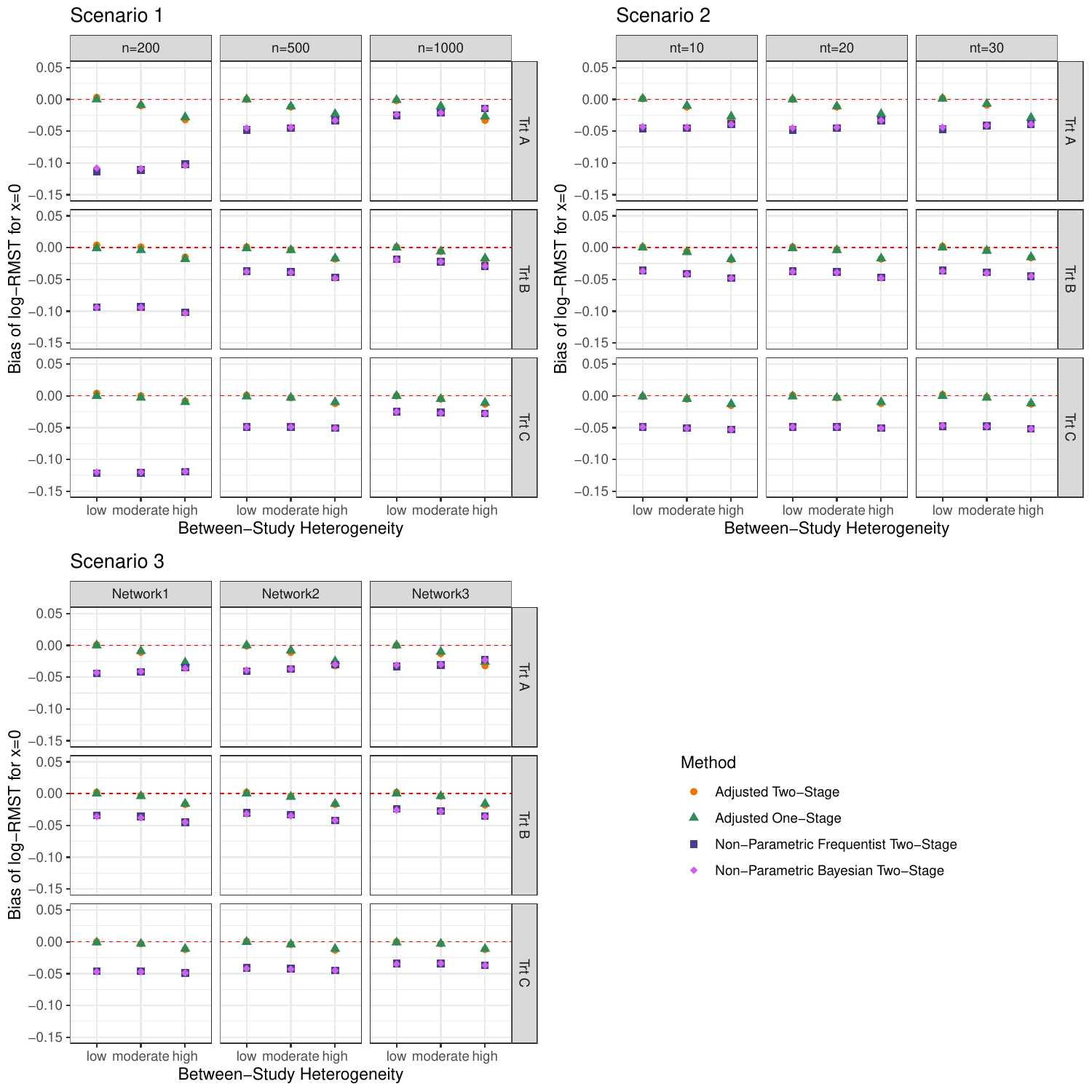}
\caption{Simulation results of bias of log-RMST of three treatments (A, B, and C) for the subgroup with $x=0$ under three scenarios.}
\label{fig:1}
\end{figure}

Figure~\ref{fig:1} presents the bias of log-RMST of three treatments for the subgroup with $x=0$ under each scenario. Compared with the existing two-stage methods, NPF and NPB, our proposed models provide no or smaller bias across all scenarios. The largest difference in bias between the existing and proposed methods is observed when $n=200$ in Scenario 1. Across all scenarios, bias tends to increase as the level of between-study heterogeneity increases. That is, NMA data containing high heterogeneity tend to yield biased estimates. When considering the same level of heterogeneity, the bias from the existing approaches decreases as the sample size in each trial increases.
However, the sample size, number of trials, and treatment network do not affect bias in our proposed methods under the same level of heterogeneity.

\begin{figure}[!htbp]
\centering\includegraphics[scale = 0.6]{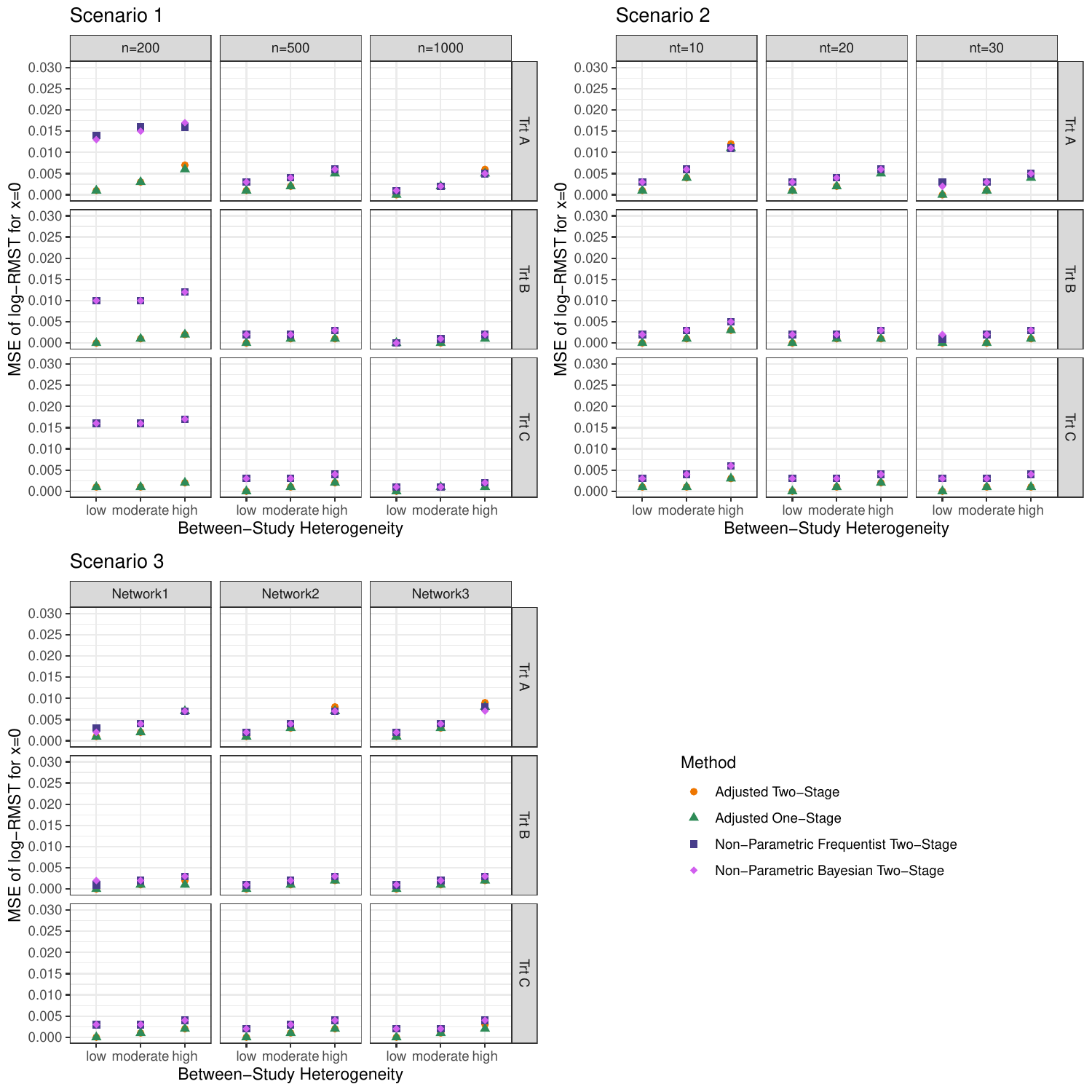}
\caption{Simulation results of mean squared error (MSE) of log-RMST of three treatments (A, B, and C) for the subgroup with $x=0$ under three scenarios.}
\label{fig:2}
\end{figure}

Figure~\ref{fig:2} displays the MSE of log-RMST for the subgroup with $x=0$. Overall, our proposed methods show MSEs no greater than those under the existing methods. In Scenario 1, the existing models provide much larger MSE than the proposed models, when $n=200$. In Scenarios 2 and 3, the MSE difference between the proposed and existing models is small. Furthermore, MSEs tend to increase as the level of between-study heterogeneity increases. In addition, fixing the level of between-study heterogeneity, all models yield decreasing MSEs when $n$ increases in Scenario 1, $nt$ increases in Scenario 2, and the network setting moves from Network 3 to Network 1 in Scenario 3. These results are expected because they show that MSEs get smaller when more data (e.g., large sample size, many trials) are used in NMA.

\begin{figure}[!htbp]
\centering\includegraphics[scale = 0.6]{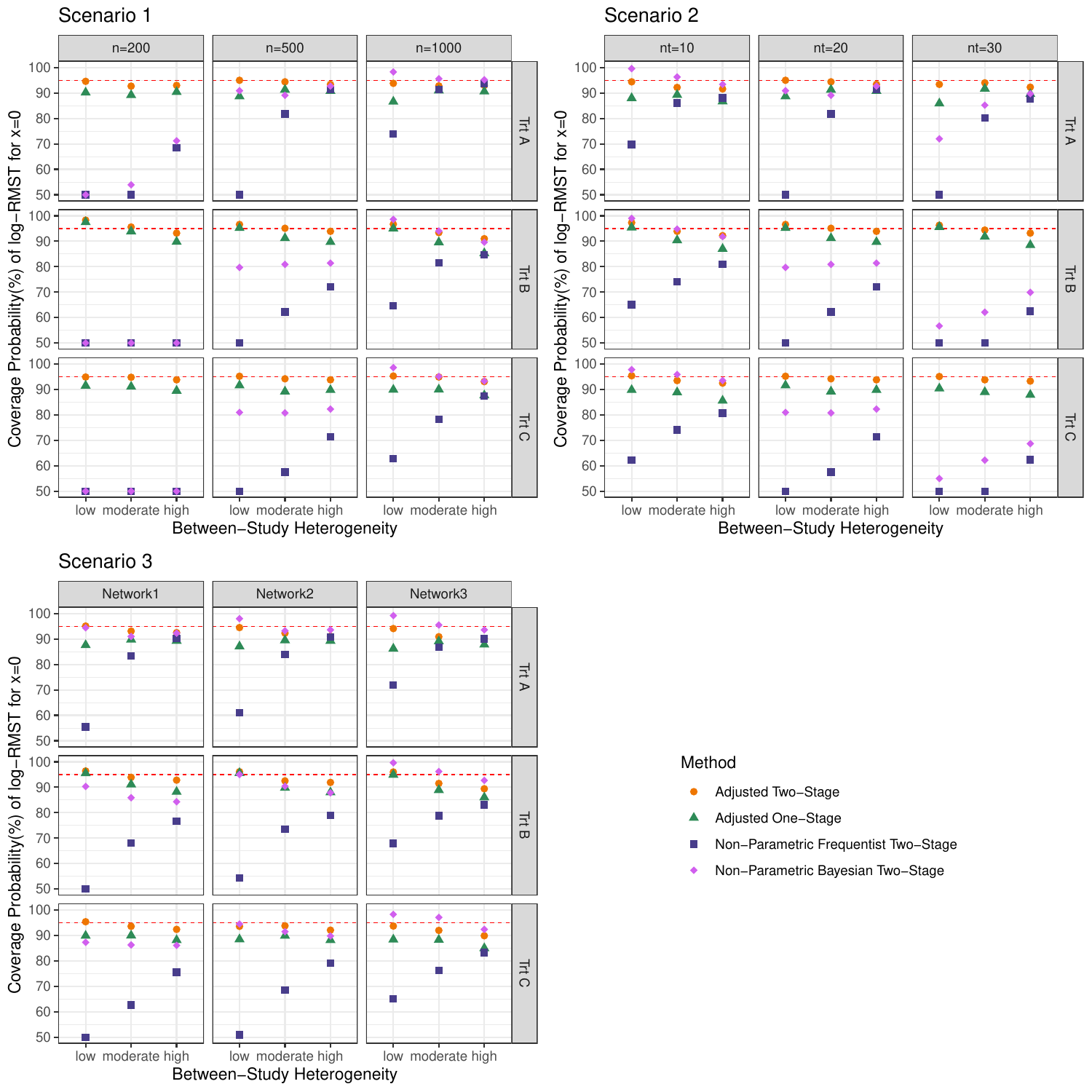}
\caption{Simulation results of coverage probabilities of log-RMST of three treatments (A, B, and C) for the subgroup with $x=0$ under three scenarios. The coverage probabilities below 50\% are truncated to 50\%.}
\label{fig:3}
\end{figure}

Figure~\ref{fig:3} presents the coverage probability of log-RMST for the subgroup with $x=0$. The existing methods provide coverage probabilities a lot lower than 95\%, especially when between-study heterogeneity is low and $n$ or $nt$ is small. In contrast, our proposed two-stage model yields coverage probabilities close to a nominal probability of 95\% in most scenarios. However, our one-stage model shows a slightly lower coverage than the two-stage model with coverage probability below 90\% in Scenario 2 when $nt=10$ and in Scenario 3 under Network 3. This shows that the standard errors of log-RMST might be underestimated in our one-stage model due to biased estimate of between-study variability (simulation results not shown). Fixing level of between-study heterogeneity, the changes in $n$, $nt$ and treatment networks have minimal effect on the coverage probabilities in our proposed methods.

In summary, the proposed RMST IPD-NMA models outperform the existing models with smaller biases, lower MSE, and coverage probabilities close to 95\% across all scenarios. Notably, the proposed models perform particularly well when overall sample sizes are small. The simulation results for subgroup $x=1$ provide similar results to those for $x=0$, which are shown in Web Figures 2 to 4 in the Supplementary Material.

\section{Data Analysis} \label{s:data analysis}
In this section, we illustrate the proposed two-stage and one-stage models using a real NMA example about treatments for patients with atrial fibrillation. The existing methods, NPF and NPB, are not applied because the data include continuous covariates that cannot be handled properly in the existing methods.

\subsection{COMBINE-AF Data}
The COMBINE-AF (A COllaboration between Multiple institutions to Better Investigate Non-vitamin K antagonist oral anticoagulant usE in Atrial Fibrillation) data contain IPD from four phase III randomized trials comparing direct oral anticoagulants (DOACs) with warfarin to prevent stroke for patients with atrial fibrillation~\citep{carnicelli2021individual,carnicelli2022direct}. Among the four trials, two are three-arm trials comparing two dose levels of DOACs, such as standard-dose DOAC (SD-DOAC) and low-dose DOAC (LD-DOAC), with warfarin, and the other two compared SD-DOAC with warfarin. The network of treatments and detailed information of individual trials can be found elsewhere~\citep{carnicelli2021individual,carnicelli2022direct}. 

In this paper, we consider two outcomes: time to bleeding and time to all-cause mortality (death) as safety and efficacy profiles, respectively. The time-to-event outcomes are in the unit of month, and we use 32 months, the point at which less than 10\% of patients remained at risk across all studies~\citep{carnicelli2022direct}, as the truncation time for RMST. We fit our proposed two-stage and one-stage models adjusted by two baseline covariates: sex as a binary variable and age as a continuous variable. The treatment-by-covariate interactions are included. We use male as the reference group and center the age to 70.55, the average age of the 4 trials.

\subsection{Results}
\begin{figure}[!htbp]
\centering\includegraphics[scale = 0.6]{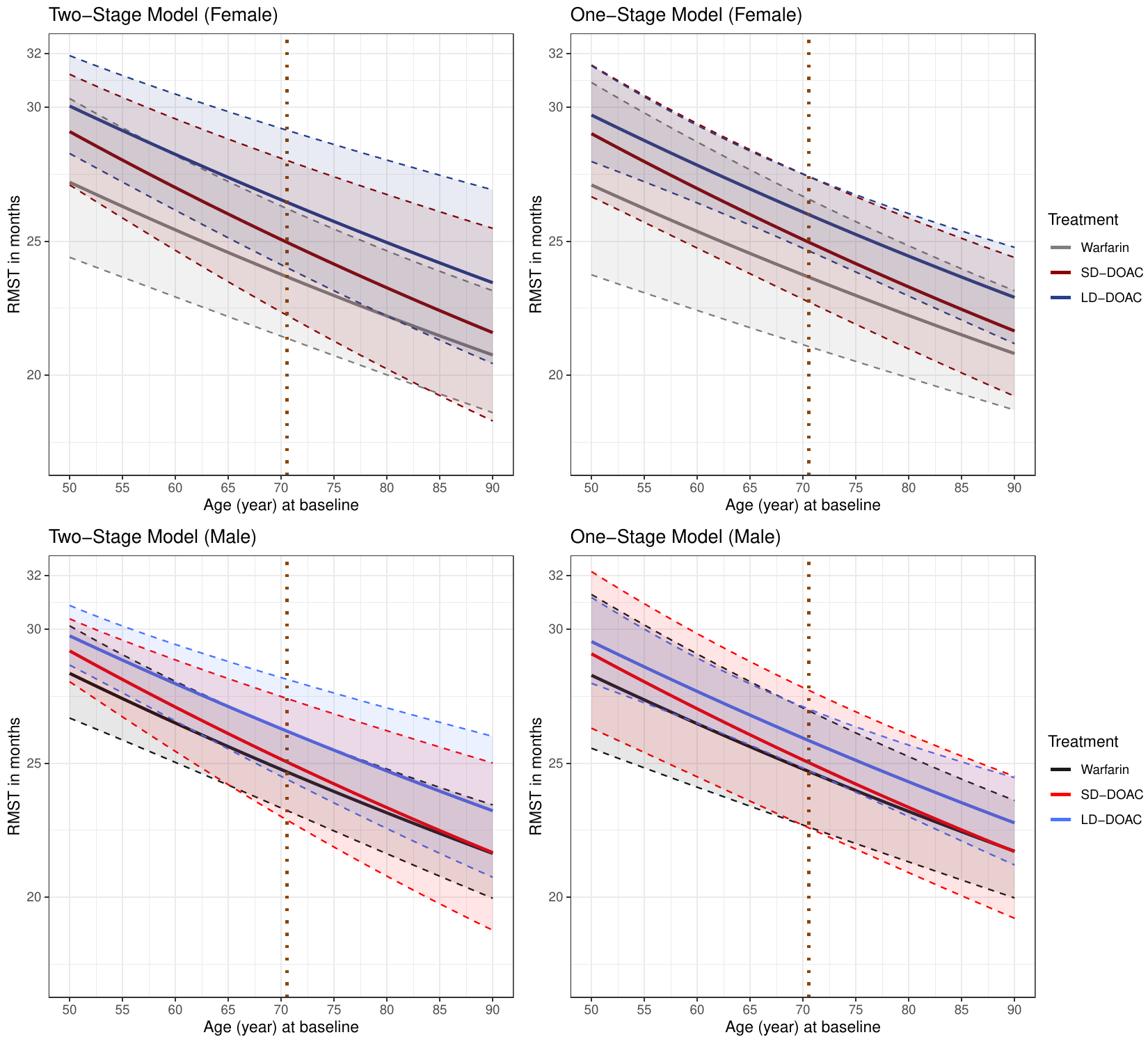}
\caption{Estimated RMSTs (in months) for the bleeding outcome over age (from 50 years to 90 years) by treatment and sex groups. The results for female and male patients are shown in the first and the second rows, respectively. The results from the two-stage and one-stage models are displayed in the first and second columns, respectively.}
\label{fig:4}
\end{figure}

\begin{figure}[!htbp]
\centering\includegraphics[scale = 0.6]{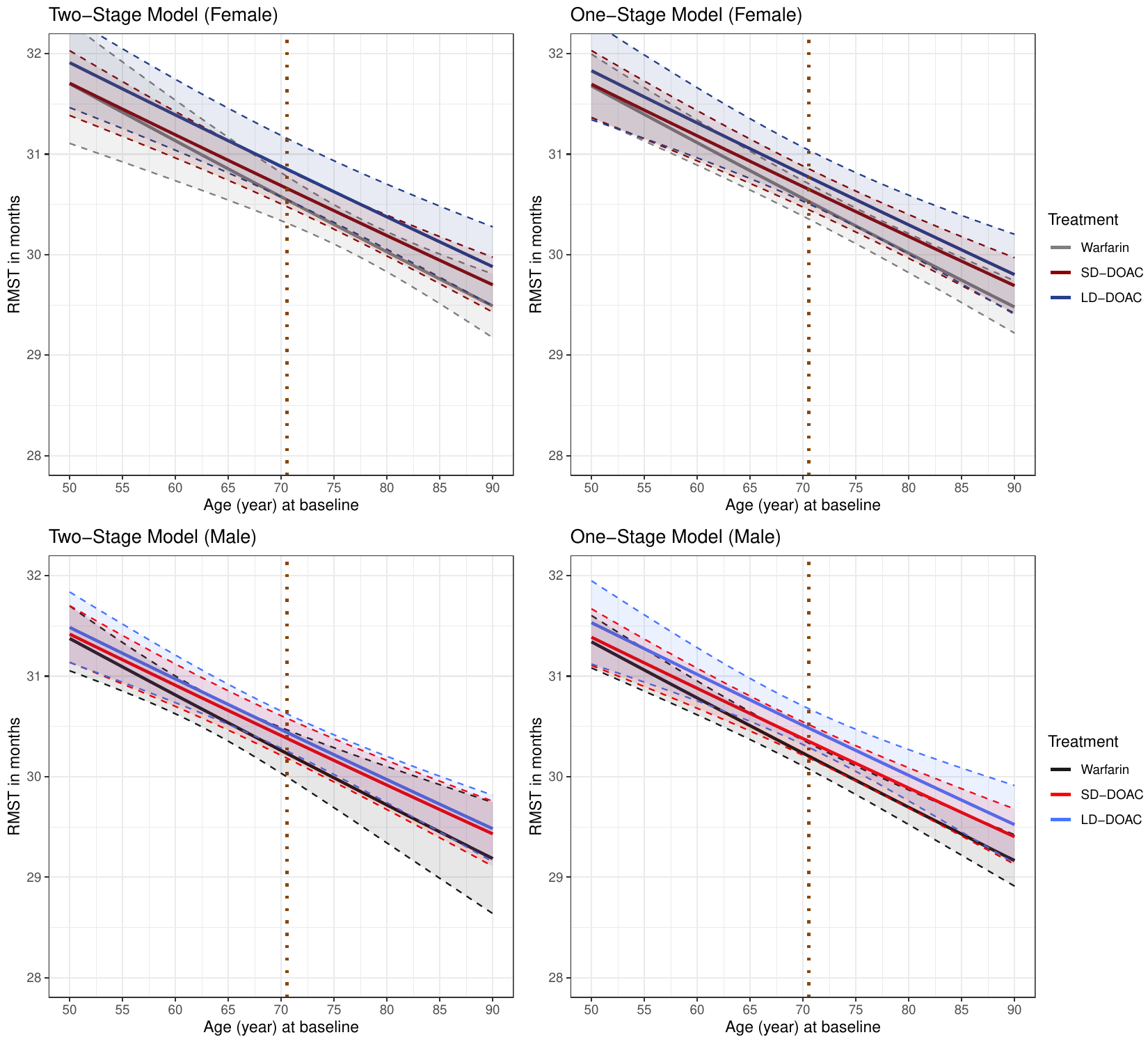}
\caption{Estimated RMSTs (months) for the all-cause mortality outcome over age (from 50 years to 90 years) by treatment and sex groups. The results for female and male patients are shown in the first and the second rows, respectively. The results from the two-stage and one-stage models are displayed in first and second columns, respectively.}
\label{fig:5}
\end{figure}

Figures~\ref{fig:4} and~\ref{fig:5} display the estimated RMSTs over age (from 50 to 90) by treatment and sex groups for the event of bleeding and all-cause mortality, respectively. The coefficient estimates of the fixed effect and the estimates of standard deviation of random effects are reported in Web Table 3 in the Supplementary Material. The proposed two-stage and one-stage models yield similar point estimates.

In Figure~\ref{fig:4}, the estimated RMSTs for the bleeding outcome demonstrate a consistent decrease as patients become more aged, and this pattern is observed across in both female and male patients and all three treatments. According to both models, all treatments have significant interaction effects with age (p-value $<$ 0.01), while the treatment effects do not significantly differ by sex (see Web Table 3 for details). LD-DOAC is preferable for avoiding bleeding among three treatments across the age range of 50 to 90 irrespective of the sex. However, we observe that the relative efficacy comparing SD-DOAC to warfarin gets smaller in senior patients.

Under the two-stage model, given patients whose age are 70.55, the estimated average bleeding-free survival times at 32 months are 26.21 months for male patients and 26.47 months for female patients randomized to LD-DOAC. The one-stage model yields corresponding results of 25.84 months for males and 25.99 months for females. Meanwhile, the two-stage model suggests a greater benefit of LD-DOAC vs. warfarin among patients at 70.55 years old. Compared to the patients randomized to warfarin, those randomized to LD-DOAC show a significant enhancement in average bleeding-free survival times at 32 months, with a relative increase of 6\% (95\% CI: [4\%, 9\%]) for males and 12\% (95\% CI: [9\%, 15\%]) for females.

Figure~\ref{fig:5} shows a similar declining pattern in the estimated RMSTs for all-cause mortality over age, which is commonly observed across sex and treatment groups. However, the estimated RMSTs are closely aligned among three treatments. Both two-stage and one-stage models reveal significant interaction effects of age with all three treatments (p-value $<$ 0.01). The two-stage model indicates that SD-DOAC and LD-DOAC have significantly different effects by sex, while the one-stage model shows that warfarin and SD-DOAC have significantly different effects by sex (see Web Table 3 for details). Under both models, LD-DOAC is preferable in mitigating the risk of death for patients whose age is between 40 and 90 regardless of sex. Additionally, as patients become more aged, SD-DOAC shows a more favorable impact on reducing all-cause mortality compared to warfarin.

Under the two-stage model, given patients whose age are 70.55, the estimated average survival times at 32 months, are 30.45 months for male patients and 30.85 months for females patients randomized to LD-DOAC. The one-stage model yields corresponding results of 30.48 months for males and 30.75 months for females. The relative RMST differences among treatments are all less than 1\% according to both models.

Between-study variability is assessed through the variance of random effects for the main treatment effects and treatment-by-covariate interactions (see Web Table 3 for details). For the bleeding outcome, the variability seems moderate for the main treatment effects and treatment-by-sex interactions, while it is notably low for treatment-by-age interactions. For the death outcome, the between-study variability for all parameters is minimal, indicating the consistency of the results across trials.

\section{Discussion} \label{s:discussion}
We developed two-stage and one-stage IPD-NMA methods that estimate RMST for time-to-event outcomes. These models used individual participant-level covariates and estimated RMST and treatment-by-covariate interaction effects using inverse probability of censoring weighted RMST regression. Our simulation study showed that the proposed two-stage and one-stage models provided unbiased RMST estimates with smaller MSE and proper coverage probability than the existing nonparametric two-stage approaches. In addition, we demonstrated the proposed methods using a real data example.

In the simulation study, we observed finite biases in both of our proposed one-stage and two-stage methods, particularly when between-study heterogeneity is high. For the two-stage approach, our findings aligned with those in several previous research~\citep{chen2012method,jackson2013matrix}, where considerable biases and low coverage probability of coefficients were observed in multivariate meta-analysis under high between-study variability. For the one-stage approach, these biases arose due to the biased PQL estimators of either the fixed effect or the standard deviations of random effects~\citep{lin1996bias,breslow2004whither,lin2007estimation,jiang2007linear,nugentbias}. The PQL algorithm involves the Laplace approximation for the integrated quasi-likelihood~\citep{breslow1993approximate}, and this approach may lead to less accurate estimations when the number of clusters (e.g., the number of studies) is small and the variance component (e.g., the between-study heterogeneity) is large~\citep{bellamy2005quantifying,jang2009numerical}. However, it is important to note that the PQL approach offers computational advantages because it can be applied to the RMST regression using the quasi-likelihood approach. It tends to perform well and these biases can be ignored when the variances of random effects are small, and the outcomes become closer to a normal distribution ~\citep{breslow2004whither,bellamy2005quantifying,jiang2007linear}. To further enhance the estimation in our one-stage method, the adaptation of alternative approximation methods, such as Gauss-Hermite quadrature can be considered~\citep{liu1994note,pinheiro1995approximations}. However, this is beyond scope of our work and remains as future research.

We used the COMBINE-AF as an illustrative example and considered selected outcomes to demonstrate the proposed methods described in the manuscript. As such, it is important to interpret the results with caution and refrain from employing them directly for clinical decision-making purposes. Clinical decisions should be made by carefully considering other important outcomes and relevant medical information. A comprehensive discussion on clinical decision-making based on the COMBINE-AF can be found in~\cite{carnicelli2022direct}.

In our data analysis, both two-stage and one-stage models produced similar coefficient estimates, but they provided slightly different estimates of the standard deviation of the between-study heterogeneity. This difference could be attributed to the varying estimation methods used in the two approaches, as well as the different likelihood functions employed in each method. In specific, the two-stage model assumes normality for all parameters derived from the first-stage, while the one-stage model uses quasi-likelihood functions.

In this paper, all NMA models were fitted under the \emph{consistency} assumption that assumes direct and indirect comparisons provide the same effects. In NMA, \emph{inconsistency} is often defined as statistical disagreement between direct and indirect comparisons~\citep{higgins2012consistency}, and it could lead to biased results if not handled properly~\citep{white2012consistency,freeman2019identifying}. However, the treatment network of our data example does not have sufficient information to assess inconsistency. For time-to-event outcomes, violations of the PH assumptions could result in inconsistency with biased indirect treatment comparisons~\citep{jansen2011network}. However, no formal method has been proposed to analyze NMA with time-to-event in the existence of inconsistency. This topic requires further research.

\section*{Software}
The relevant R code for the methodology and simulation study is available on \url{https://github.com/kimihua1995/NMA_IPD_RMST}.

\section*{Supplementary Material}
Web Appendices, Tables and Figures, referenced in Sections~\ref{s:methods} to~\ref{s:data analysis} are available at the Supplementary Material, which contains 1) a brief summary of the algorithm for fitting the one-stage RMST IPD-NMA model (Equation~\eqref{eqn:3}); 2) additional simulation settings and results; 3) additional data analysis results.

\section*{Data Availability Statement}
The data that support the findings of this study are available to members of the COMBINE-AF executive committee. Restrictions apply to the availability of these data, which were used under license for this study. Individual investigators may reach out directly to the COMBINE-AF executive committee for collaboration.

\section*{Acknowledgments}
Dr. Hong was partially supported by the National Institute of Mental Health (R00MH111807 and R01MH126856) and Dr. Wang was partially supported by the NCI (P01 CA142538), the NIA (R01 AG066883), and the FDA (U01 FD007934).
{\it Conflict of Interest}: None declared.

\printbibliography

@article{dersimonian1986meta,
  title={Meta-analysis in clinical trials},
  author={DerSimonian,R and Laird, N},
  journal={Controlled Clinical Trials},
  volume={7},
  number={3},
  pages={177--188},
  year={1986},
  publisher={Elsevier}
}

@article{hong2016bayesian,
  title={A {B}ayesian missing data framework for generalized multiple outcome mixed treatment comparisons},
  author={Hong, Hwanhee and Chu, Haitao and Zhang, Jing and Carlin, Bradley P},
  journal={Research Synthesis Methods},
  volume={7},
  number={1},
  pages={6--22},
  year={2016},
  publisher={Wiley Online Library}
}

@article{white2015network,
  title={Network meta-analysis},
  author={White, Ian R},
  journal={The Stata Journal},
  volume={15},
  number={4},
  pages={951--985},
  year={2015},
  publisher={SAGE Publications Sage CA: Los Angeles, CA}
}

@article{salanti2008evaluation,
  title={Evaluation of networks of randomized trials},
  author={Salanti, Georgia and Higgins, Julian PT and Ades, AE and Ioannidis, John PA},
  journal={Statistical Methods in Medical Research},
  volume={17},
  number={3},
  pages={279--301},
  year={2008},
  publisher={SAGE Publications Sage UK: London, England}
}

@article{mills2013demystifying,
  title={Demystifying trial networks and network meta-analysis},
  author={Mills, Edward J and Thorlund, Kristian and Ioannidis, John PA},
  journal={BMJ},
  volume={346},
  year={2013},
  publisher={British Medical Journal Publishing Group}
}

@article{Saramago14,
  title={Network meta-analysis of (individual patient) time to event data alongside (aggregate) count data},
  author={Saramago, Pedro and Chuang, Ling-Hsiang and Soares, Marta O},
  journal={BMC Medical Research Methodology},
  volume={14},
  number={1},
  pages={1--11},
  year={2014},
  publisher={BioMed Central}
}

@article{lu2004combination,
  title={Combination of direct and indirect evidence in mixed treatment comparisons},
  author={Lu, Guobing and Ades, A E},
  journal={Statistics in Medicine},
  volume={23},
  number={20},
  pages={3105--3124},
  year={2004},
  publisher={Wiley Online Library}
}

@article{donegan2013combining,
  title={Combining individual patient data and aggregate data in mixed treatment comparison meta-analysis: individual patient data may be beneficial if only for a subset of trials},
  author={Donegan, Sarah and Williamson, Paula and D'Alessandro, Umberto and Garner, Paul and Smith, Catrin Tudur},
  journal={Statistics in Medicine},
  volume={32},
  number={6},
  pages={914--930},
  year={2013},
  publisher={Wiley Online Library}
}

@article{donegan2012assessing,
  title={Assessing the consistency assumption by exploring treatment by covariate interactions in mixed treatment comparison meta-analysis: individual patient-level covariates versus aggregate trial-level covariates},
  author={Donegan, Sarah and Williamson, Paula and D'Alessandro, Umberto and Smith, Catrin Tudur},
  journal={Statistics in Medicine},
  volume={31},
  number={29},
  pages={3840--3857},
  year={2012},
  publisher={Wiley Online Library}
}

@article{Hong15,
  title={Incorporation of individual-patient data in network meta-analysis for multiple continuous endpoints, with application to diabetes treatment},
  author={Hong, Hwanhee and Fu, Haoda and Price, Karen L and Carlin, Bradley P},
  journal={Statistics in Medicine},
  volume={34},
  number={20},
  pages={2794--2819},
  year={2015},
  publisher={Wiley Online Library}
}

@article{debray2015get,
  title={Get real in individual participant data (IPD) meta-analysis: a review of the methodology},
  author={Debray, Thomas P A and Moons, Karel GM and {van Valkenhoef}, Gert and Efthimiou, Orestis and Hummel, Noemi and Groenwold, Rolf HH and Reitsma, Johannes B and GetReal Methods Review Group},
  journal={Research Synthesis Methods},
  volume={6},
  number={4},
  pages={293--309},
  year={2015},
  publisher={Wiley Online Library}
}

@article{lyman2005strengths,
  title={The strengths and limitations of meta-analyses based on aggregate data},
  author={Lyman, Gary H and Kuderer, Nicole M},
  journal={BMC Medical Research Methodology},
  volume={5},
  number={1},
  pages={1--7},
  year={2005},
  publisher={BioMed Central}
}

@article{riley2010meta,
  title={Meta-analysis of individual participant data: rationale, conduct, and reporting},
  author={Riley, Richard D and Lambert, Paul C and Abo-Zaid, Ghada},
  journal={BMJ : British Medical Journal},
  volume={340},
  year={2010},
  publisher={British Medical Journal Publishing Group}
}

@article{simmonds2005meta,
  title={Meta-analysis of individual patient data from randomized trials: a review of methods used in practice},
  author={Simmonds, Mark C and Higginsa, Julian PT and Stewartb, Lesley A and Tierneyb, Jayne F and Clarke, Mike J and Thompson, Simon G},
  journal={Clinical Trials},
  volume={2},
  number={3},
  pages={209--217},
  year={2005},
  publisher={Sage Publications Sage CA: Thousand Oaks, CA}
}

@article{stewart2012statistical,
  title={Statistical analysis of individual participant data meta-analyses: a comparison of methods and recommendations for practice},
  author={Stewart, Gavin B and Altman, Douglas G and Askie, Lisa M and Duley, Lelia and Simmonds, Mark C and Stewart, Lesley A},
  year={2012},
  publisher={Public Library of Science San Francisco, USA}
}

@article{kontopantelis2018comparison,
  title={A comparison of one-stage vs two-stage individual patient data meta-analysis methods: A simulation study},
  author={Kontopantelis, Evangelos},
  journal={Research Synthesis Methods},
  volume={9},
  number={3},
  pages={417--430},
  year={2018},
  publisher={Wiley Online Library}
}

@article{hua2017one,
  title={One-stage individual participant data meta-analysis models: estimation of treatment-covariate interactions must avoid ecological bias by separating out within-trial and across-trial information},
  author={Hua, Hairui and Burke, Danielle L and Crowther, Michael J and Ensor, Joie and Tudur Smith, Catrin and Riley, Richard D},
  journal={Statistics in Medicine},
  volume={36},
  number={5},
  pages={772--789},
  year={2017},
  publisher={Wiley Online Library}
}

@article{Smith05,
  title={Investigating heterogeneity in an individual patient data meta-analysis of time to event outcomes},
  author={Smith, Catrin Tudur and Williamson, Paula R and Marson, Anthony G},
  journal={Statistics in Medicine},
  volume={24},
  number={9},
  pages={1307--1319},
  year={2005},
  publisher={Wiley Online Library}
}

@article{bowden2011individual,
  title={Individual patient data meta-analysis of time-to-event outcomes: one-stage versus two-stage approaches for estimating the hazard ratio under a random effects model},
  author={Bowden, Jack and Tierney, Jayne F and Simmonds, Mark and Copas, Andrew J and Higgins, Julian PT},
  journal={Research Synthesis Methods},
  volume={2},
  number={3},
  pages={150--162},
  year={2011},
  publisher={Wiley Online Library}
}

@article{cox1972regression,
  title={Regression models and life-tables},
  author={Cox, D.R.},
  journal={Journal of the Royal Statistical Society: Series B (Methodological)},
  volume={34},
  number={2},
  pages={187--202},
  year={1972},
  publisher={Wiley Online Library}
}

@article{crowther2012individual,
  title={Individual patient data meta-analysis of survival data using {P}oisson regression models},
  author={Crowther, Michael J and Riley, Richard D and Staessen, Jan A and Wang, Jiguang and Gueyffier, Francois and Lambert, Paul C},
  journal={BMC Medical Research Methodology},
  volume={12},
  number={1},
  pages={1--14},
  year={2012},
  publisher={BioMed Central}
}

@article{jansen2011network,
  title={Network meta-analysis of survival data with fractional polynomials},
  author={Jansen, Jeroen P},
  journal={BMC Medical Research Methodology},
  volume={11},
  number={1},
  pages={1--14},
  year={2011},
  publisher={BioMed Central}
}

@article{irwin1949standard,
  title={The standard error of an estimate of expectation of life, with special reference to expectation of tumourless life in experiments with mice},
  author={Irwin, J.O.},
  journal={Epidemiology \& Infection},
  volume={47},
  number={2},
  pages={188--189},
  year={1949},
  publisher={Cambridge University Press}
}

@article{lin1989robust,
  title={The robust inference for the Cox proportional hazards model},
  author={Lin, D. and Wei, L.},
  journal={Journal of the American Statistical Association},
  volume={84},
  number={408},
  pages={1074--1078},
  year={1989},
  publisher={Taylor \& Francis}
}

@article{royston2013restricted,
  title={Restricted mean survival time: an alternative to the hazard ratio for the design and analysis of randomized trials with a time-to-event outcome},
  author={Royston, P. and Parmar, M. KB.},
  journal={BMC Medical Research Methodology},
  volume={13},
  number={1},
  pages={1--15},
  year={2013},
  publisher={BioMed Central}
}

@article{andersen2004regression,
  title={Regression analysis of restricted mean survival time based on pseudo-observations},
  author={Andersen, P.K. and Hansen, M.G. and Klein, J.P.},
  journal={Lifetime Data Analysis},
  volume={10},
  number={4},
  pages={335--350},
  year={2004},
  publisher={Springer}
}

@article{tian2014predicting,
  title={Predicting the restricted mean event time with the subject's baseline covariates in survival analysis},
  author={Tian, L. and Zhao, L. and Wei, L.},
  journal={Biostatistics},
  volume={15},
  number={2},
  pages={222--233},
  year={2014},
  publisher={Oxford University Press}
}

@article{gardiner2021restricted,
  title={Restricted Mean Survival Time Estimation: Nonparametric and Regression Methods},
  author={Gardiner, J.C.},
  journal={Journal of Statistical Theory and Practice},
  volume={15},
  number={1},
  pages={1--15},
  year={2021},
  publisher={Springer}
}

@article{uno2014moving,
  title={Moving beyond the hazard ratio in quantifying the between-group difference in survival analysis},
  author={Uno, H. and Claggett, Brian and Tian, Lu and Inoue, Eisuke and Gallo, Paul and Miyata, Toshio and Schrag, Deborah and Takeuchi, Masahiro and Uyama, Yoshiaki and Zhao, Lihui and others},
  journal={Journal of Clinical Oncology},
  volume={32},
  number={22},
  pages={2380},
  year={2014},
  publisher={American Society of Clinical Oncology}
}

@article{royston2002flexible,
  title={Flexible parametric proportional-hazards and proportional-odds models for censored survival data, with application to prognostic modelling and estimation of treatment effects},
  author={Royston, P. and Parmar, M.KB.},
  journal={Statistics in Medicine},
  volume={21},
  number={15},
  pages={2175--2197},
  year={2002},
  publisher={Wiley Online Library}
}

@article{kim2017restricted,
  title={Restricted mean survival time as a measure to interpret clinical trial results},
  author={Kim, D. and Uno, H. and Wei, L.},
  journal={JAMA Cardiology},
  volume={2},
  number={11},
  pages={1179--1180},
  year={2017},
  publisher={American Medical Association}
}

@article{perego2020utility,
  title={Utility of restricted mean survival time analysis for heart failure clinical trial evaluation and interpretation},
  author={Perego, C. and Sbolli, Marco and Specchia, Claudia and Fiuzat, Mona and McCaw, Zachary R and Metra, Marco and Oriecuia, Chiara and Peveri, Giulia and Wei, Lee-Jen and O’Connor, Christopher M and others},
  journal={Heart Failure},
  volume={8},
  number={12},
  pages={973--983},
  year={2020},
  publisher={American College of Cardiology Foundation Washington DC}
}

@article{wei2015meta,
  title={Meta-analysis of time-to-event outcomes from randomized trials using restricted mean survival time: application to individual participant data},
  author={Wei, Y. and Royston, P. and Tierney, J.F. and Parmar, M.KB.},
  journal={Statistics in Medicine},
  volume={34},
  number={21},
  pages={2881--2898},
  year={2015},
  publisher={Wiley Online Library}
}

@article{lueza2016difference,
  title={Difference in restricted mean survival time for cost-effectiveness analysis using individual patient data meta-analysis: evidence from a case study},
  author={Lueza, B. and Mauguen, A. and Pignon, Jean-Pierre and Rivero-Arias, Oliver and Bonastre, Julia and MAR-LC Collaborative Group},
  journal={PLoS One},
  volume={11},
  number={3},
  pages={e0150032},
  year={2016},
  publisher={Public Library of Science San Francisco, CA USA}
}

@article{weir2021multivariate,
  title={Multivariate meta-analysis model for the difference in restricted mean survival times},
  author={Weir, I.R. and Tian, L. and Trinquart, L.},
  journal={Biostatistics},
  volume={22},
  number={1},
  pages={82--96},
  year={2021},
  publisher={Oxford University Press}
}

@article{tang2022bayesian,
  title={Bayesian multivariate network meta-analysis model for the difference in restricted mean survival times},
  author={Tang, X. and Trinquart, L.},
  journal={Statistics in Medicine},
  volume={41},
  number={3},
  pages={595--611},
  year={2022},
  publisher={Wiley Online Library}
}

@article{burke2017meta,
  title={Meta-analysis using individual participant data: one-stage and two-stage approaches, and why they may differ},
  author={Burke, D.L. and Ensor, J. and Riley, R.D.},
  journal={Statistics in Medicine},
  volume={36},
  number={5},
  pages={855--875},
  year={2017},
  publisher={Wiley Online Library}
}

@article{hawkins2016arm,
  title={‘Arm-based’parameterization for network meta-analysis},
  author={Hawkins, Neil and Scott, David A and Woods, Beth},
  journal={Research Synthesis Methods},
  volume={7},
  number={3},
  pages={306--313},
  year={2016},
  publisher={Wiley Online Library}
}

@article{debray2013individual,
  title={Individual participant data meta-analysis for a binary outcome: one-stage or two-stage?},
  author={Debray, Thomas PA and Moons, Karel GM and Abo-Zaid, Ghada Mohammed Abdallah and Koffijberg, Hendrik and Riley, Richard David},
  journal={PloS One},
  volume={8},
  number={4},
  pages={e60650},
  year={2013},
  publisher={Public Library of Science San Francisco, USA}
}

@article{carnicelli2021individual,
  title={Individual Patient Data from the Pivotal Randomized Controlled Trials of Non-Vitamin K Antagonist Oral Anticoagulants in Patients with Atrial Fibrillation (COMBINE AF): Design and Rationale: From the COMBINE AF (A Collaboration between Multiple institutions to Better Investigate Non-vitamin K antagonist oral anticoagulant use in Atrial Fibrillation) Investigators},
  author={Carnicelli, A.P. and Hong, Hwanhee and Giugliano, Robert P and Connolly, Stuart J and Eikelboom, John and Patel, Manesh R and Wallentin, Lars and Morrow, David A and Wojdyla, Daniel and Hua, Kaiyuan and others},
  journal={American Heart Journal},
  volume={233},
  pages={48--58},
  year={2021},
  publisher={Elsevier}
}

@article{carnicelli2022direct,
  title={Direct oral anticoagulants versus warfarin in patients with atrial fibrillation: patient-level network meta-analyses of randomized clinical trials with interaction testing by age and sex},
  author={Carnicelli, A.P. and Hong, Hwanhee and Connolly, Stuart J and Eikelboom, John and Giugliano, Robert P and Morrow, David A and Patel, Manesh R and Wallentin, Lars and Alexander, John H and Bahit, M Cecilia and others},
  journal={Circulation},
  year={2022},
  publisher={Am Heart Assoc}
}

@article{breslow1993approximate,
  title={Approximate inference in generalized linear mixed models},
  author={Breslow, Norman E and Clayton, David G},
  journal={Journal of the American statistical Association},
  volume={88},
  number={421},
  pages={9--25},
  year={1993},
  publisher={Taylor \& Francis}
}

@article{lin1996bias,
  title={Bias correction in generalized linear mixed models with multiple components of dispersion},
  author={Lin, Xihong and Breslow, Norman E},
  journal={Journal of the American Statistical Association},
  volume={91},
  number={435},
  pages={1007--1016},
  year={1996},
  publisher={Taylor \& Francis}
}

@article{lin2007estimation,
  title={Estimation using penalized quasilikelihood and quasi-pseudo-likelihood in {P}oisson mixed models},
  author={Lin, Xihong},
  journal={Lifetime Data Analysis},
  volume={13},
  pages={533--544},
  year={2007},
  publisher={Springer}
}

@article{jackson2013matrix,
  title={A matrix-based method of moments for fitting the multivariate random effects model for meta-analysis and meta-regression},
  author={Jackson, D. and White, I.R. and Riley, R.D.},
  journal={Biometrical Journal},
  volume={55},
  number={2},
  pages={231--245},
  year={2013},
  publisher={Wiley Online Library}
}

@article{jackson2011multivariate,
  title={Multivariate meta-analysis: potential and promise},
  author={Jackson, Dan and Riley, Richard and White, Ian R},
  journal={Statistics in medicine},
  volume={30},
  number={20},
  pages={2481--2498},
  year={2011},
  publisher={Wiley Online Library}
}

@article{wei2013bayesian,
  title={Bayesian multivariate meta-analysis with multiple outcomes},
  author={Wei, Yinghui and Higgins, Julian PT},
  journal={Statistics in Medicine},
  volume={32},
  number={17},
  pages={2911--2934},
  year={2013},
  publisher={Wiley Online Library}
}

@article{chen2012method,
  title={A method of moments estimator for random effect multivariate meta-analysis},
  author={Chen, H. and Manning, A.K. and Dupuis, J.},
  journal={Biometrics},
  volume={68},
  number={4},
  pages={1278--1284},
  year={2012},
  publisher={Wiley Online Library}
}

@book{schwarzer2015meta,
  title={Meta-analysis with R},
  author={Schwarzer, Guido and Carpenter, James R and R{\"u}cker, Gerta and others},
  volume={4784},
  year={2015},
  publisher={Springer}
}

@article{higgins2012consistency,
  title={Consistency and inconsistency in network meta-analysis: concepts and models for multi-arm studies},
  author={Higgins, JPT and Jackson, D and Barrett, JK and Lu, G and Ades, AE and White, IR},
  journal={Research Synthesis Methods},
  volume={3},
  number={2},
  pages={98--110},
  year={2012},
  publisher={Wiley Online Library}
}

@article{freeman2019identifying,
  title={Identifying inconsistency in network meta-analysis: Is the net heat plot a reliable method?},
  author={Freeman, Suzanne C and Fisher, David and White, Ian R and Auperin, Anne and Carpenter, James R},
  journal={Statistics in Medicine},
  volume={38},
  number={29},
  pages={5547--5564},
  year={2019},
  publisher={Wiley Online Library}
}

@article{white2019comparison,
  title={A comparison of arm-based and contrast-based models for network meta-analysis},
  author={White, Ian R and Turner, Rebecca M and Karahalios, Amalia and Salanti, Georgia},
  journal={Statistics in Medicine},
  volume={38},
  number={27},
  pages={5197--5213},
  year={2019},
  publisher={Wiley Online Library}
}

@article{white2012consistency,
  title={Consistency and inconsistency in network meta-analysis: model estimation using multivariate meta-regression},
  author={White, Ian R and Barrett, Jessica K and Jackson, Dan and Higgins, Julian PT},
  journal={Research Synthesis Methods},
  volume={3},
  number={2},
  pages={111--125},
  year={2012},
  publisher={Wiley Online Library}
}

@article{gasparrini2015package,
  title={Package ‘mvmeta’},
  author={Gasparrini, Antonio and Gasparrini, Maintainer Antonio},
  year={2015}
}

@article{ripley2013package,
  title={Package ‘mass’},
  author={Ripley, Brian and Venables, Bill and Bates, Douglas M and Hornik, Kurt and Gebhardt, Albrecht and Firth, David and Ripley, Maintainer Brian},
  journal={Cran R},
  volume={538},
  pages={113--120},
  year={2013}
}

@book{mccullagh2019generalized,
  title={Generalized linear models},
  author={McCullagh, Peter},
  year={2019},
  publisher={Routledge}
}

@article{wang2018modeling,
  title={Modeling restricted mean survival time under general censoring mechanisms},
  author={Wang, Xin and Schaubel, Douglas E},
  journal={Lifetime data analysis},
  volume={24},
  pages={176--199},
  year={2018},
  publisher={Springer}
}

@article{zhong2022restricted,
  title={Restricted mean survival time as a function of restriction time},
  author={Zhong, Yingchao and Schaubel, Douglas E},
  journal={Biometrics},
  volume={78},
  number={1},
  pages={192--201},
  year={2022},
  publisher={Wiley Online Library}
}

@inproceedings{breslow2004whither,
  title={Whither pql?},
  author={Breslow, Norman},
  booktitle={Proceedings of the second seattle symposium in biostatistics: analysis of correlated data},
  pages={1--22},
  year={2004},
  organization={Springer}
}

@article{jang2009numerical,
  title={A numerical study of PQL estimation biases in generalized linear mixed models under heterogeneity of random effects},
  author={Jang, Woncheol and Lim, Johan},
  journal={Communications in Statistics-Simulation and Computation},
  volume={38},
  number={4},
  pages={692--702},
  year={2009},
  publisher={Taylor \& Francis}
}

@book{jiang2007linear,
  title={Linear and generalized linear mixed models and their applications},
  author={Jiang, Jiming and Nguyen, Thuan},
  volume={1},
  year={2007},
  publisher={Springer}
}

@article{bellamy2005quantifying,
  title={Quantifying PQL bias in estimating cluster-level covariate effects in generalized linear mixed models for group-randomized trials},
  author={Bellamy, Scarlett L and Li, Yi and Lin, Xihong and Ryan, Louise M},
  journal={Statistica Sinica},
  pages={1015--1032},
  year={2005},
  publisher={JSTOR}
}

@article{nugentbias,
  title={Bias induced by fitting GLMMs with dichotomous outcomes using penalized quasi-likelihood},
  author={Nugent, Joshua and Doone, Bianca and Kleinman, Ken}
}

@article{liu1994note,
  title={A note on Gauss—Hermite quadrature},
  author={Liu, Qing and Pierce, Donald A},
  journal={Biometrika},
  volume={81},
  number={3},
  pages={624--629},
  year={1994},
  publisher={Oxford University Press}
}

@article{pinheiro1995approximations,
  title={Approximations to the log-likelihood function in the nonlinear mixed-effects model},
  author={Pinheiro, Jos{\'e} C and Bates, Douglas M},
  journal={Journal of computational and Graphical Statistics},
  volume={4},
  number={1},
  pages={12--35},
  year={1995},
  publisher={Taylor \& Francis}
}

\clearpage
\begin{table}[!htbp]
\caption{Treatment networks in Scenario 3 from the simulation study. The upper table summarizes the number of trials that with each treatment combination. The lower table summaries the total number of trials that contain certain treatment.}
\label{tab:1}
\begin{center}
\begin{tabular}{lccc}
\hline
& \multicolumn{3}{c}{Number of Trials}\\
Treatments&Network 1&Network 2&Network 3\\
\hline
A,B,C & 14 & 8 & 0 \\
A,B & 2 & 4 & 7 \\
A,C & 2 & 4 & 7 \\
B,C & 2 & 4 & 6 \\
\hline
& \multicolumn{3}{c}{Total Number of Trials Studying} \\
& \multicolumn{3}{c}{Individual Treatment} \\
Treatment&Network 1&Network 2&Network 3\\
\hline
A & 18 & 16 & 14 \\
B & 18 & 16 & 13 \\
C & 18 & 16 & 13 \\
\hline
\end{tabular}
\end{center}
\end{table}

\clearpage

\clearpage

\clearpage

\clearpage

\clearpage

\end{document}


\begin{center}
{\Large{ \textbf{Supplementary Material for ``Network Meta-Analysis of Time-to-Event Endpoints with Individual Participant Data using Restricted Mean Survival Time Regression"}}} 
\bigskip
\bigskip

\textbf{Kaiyuan Hua$^{1}$, Xiaofei Wang$^{1}$, Hwanhee Hong$^{1}$}\\
$^{1}$ Department of Biostatistics and Bioinformatics, Duke University School of Medicine, Durham, North Carolina, 27705, USA \\

\end{center}

\bigskip

This supplementary material is organized as follows. Web Appendix A briefly shows the algorithm for fitting the one-stage RMST IPD-NMA model (Equation 2.3). Web Appendix B provides the simulation settings including the prespecified values of parameters in Equation 3.1 (Web Table 1) and the discussion about calculating the true estimands. Web Figure 1 shows the survival curves used in the simulation study. Web Appendix C presents the additional simulation results of bias, mean squared error and coverage probability of RMSTs for the subgroup with $x=1$ under three scenarios in Web Figures 2 to 4, respectively. In Web Appendix D, Web Table 3 presents additional data analysis results with the outcome of bleeding (upper table) and the outcome of all-cause mortality (lower table) under our proposed two-stage and one-stage methods.

\newpage
\section*{Web Appendix A. PQL Estimation for the One-Stage RMST IPD-NMA Model}
Consider the total sample size of $N = \sum_{j=1}^{J}n_j$, Equation (2.3) can be written in a matrix form as follows:
\begin{eqnarray}
   g(\bmu^{(\br)}) &=& (\bA , \bX_1, \dots, \bX_P) (\balpha^T, \bbeta_1^T, \dots, \bbeta_P^T)^T + \bZ_{A1} \ba_1' + \dots + \bZ_{AK} \ba_K' \nonumber \\ 
   &+& \sum_{l=1}^P [ \bZ_{X1p}\bb_{1p}' + \dots + \bZ_{XKp}\bb_{Kp}' ] \nonumber \\
   &=& \Tilde{\bX} \bgamma + \sum_{k=1}^{K} \bZ_{Ak} \ba_k' + \sum_{l=1}^P\sum_{k=1}^{K} \bZ_{Xkp} \bb_{kp}' \nonumber \\
   &=& \Tilde{\bX} \bgamma + \Tilde{\bZ} \br
\end{eqnarray}
In Equation A.1, $\bmu^{(\br)} = (\mu_i^{(\br)})_{i=1,\dots,N}$ is a $N \times 1$ vector, where $\mu_i^{(\br)} = E(Y_{i}|\br)$ is the conditional expectation of the response $Y_{i}$. The conditional variance for $Y_i$ is $\mbox{Var}(Y_i|\br) = \phi v(\mu_i^{(\br)})/w_i$, where $w_i$ is the inverse probability of censoring weight of the $i^{th}$ patient. $\balpha = (\alpha_1,\dots,\alpha_k)^T$ and $\bbeta_p = (\beta_{1p},\dots,\beta_{Kp})^T$ are the parameter vectors for the fixed effect. $\bA$ is a $N \times K$ design matrix for the treatments and $\bX_p$ is the $N \times K$ design matrix for the treatment-by-the $p^{th}$ covariate, for $p = 1,\dots,P$. We let $\Tilde{\bX}$ be partitioned as $(\bA , \bX_1, \dots, \bX_P)$ and $\bgamma$ be partitioned as $(\balpha^T, \bbeta_1^T, \dots, \bbeta_L^T)^T$. $\bZ_{Ak}$ is a $N\times J$ design matrix for the random treatment effects $\ba_k' = (a_{1k},\dots,a_{Jk})^T$, and $\bZ_{Xkp}$ is the $N\times J$ design matrix for the random treatment-by-the $p^{th}$ covariate effects $\bb_{kp}' = (b_{1kp},\dots,b_{Jkp})^T$. The elements of the matrices $\bZ_{Ak}$ and $\bZ_{Xkp}$ are 0 or 1. We let $\Tilde{\bZ}$ be partitioned as $ (\bZ_{A1},\dots,\bZ_{AK},\bZ_{X11},\dots,\bZ_{XK1},\dots,\bZ_{X1P}\dots,\bZ_{XKP})$ and $\br$ be partitioned as $(\ba_1',\dots,\ba_K',\dots,\bb_{11}',\dots,\bb_
{K1}',\dots,\bb_{1P}',\dots,\bb_{KP}')^T$. We assume the variance-covariance matrix of the random effects vector $\br$ is represented as $\bD$ and we refer the parameters contained in $\bD$ as $\btau$. Note that $\btau$ is a composite of parameters that are involved in $\bR_1$, which is defined in Section 2.

In the proposed one-stage model, we use the penalized quasi-likelihood (PQL) method [Breslow and Clayton, 1993] to estimate the parameters $\bgamma = (\balpha^T, \bbeta_1^T, \dots, \bbeta_L^T)^T$ for the fixed effect and $\btau$ for the random effects. The objective function for estimating the parameters $(\bgamma,\btau)$ is the integrated quasi-likelihood given by
\begin{eqnarray}
    L(\bgamma,\btau) \propto |\bR_1|^{-\frac{1}{2}}\int exp \left[ -\frac{1}{2\phi}\sum_{i=1}^{N}d_i(y_i,\mu_i^{(\br)}) - \frac{1}{2}\br^T\bD^{-1}\br \right] d\br,
\end{eqnarray}
where
\[
d_i(y,\mu) = -2w_i\int_y^\mu \frac{y - u}{v(u)}du.
\]
Since the integral from Equation (A.2) cannot be evaluated as a closed form, the Laplace's method is applied for the integral approximation. As such, maximizing Equation (A.2) is approximated to maximize
\begin{eqnarray}
    \mbox{PQL}(\bgamma,\br) =  -\frac{1}{2\phi}\sum_{i=1}^{N}d_i(y_i,\mu_i^{(\br)}) - \frac{1}{2}\br^T\bD^{-1}\br,
\end{eqnarray}
which is the \emph{penalized quasi-likelihood (PQL)}. Here, the components of $\br$ serve as predictors of the random effects, and $\bgamma = \bgamma(\btau)$ and $\br = \br(\btau)$ are both functions of $\btau$. For fixing $\btau$, differentiation of Equation (A.3) with respect to $\bgamma$ and $\br$ leads to the following score functions:
\begin{eqnarray}
    \sum_{i=1}^N \frac{(y_i - \mu_i^{(\br)})\Tilde{X}_i}{\phi w_i v(\mu_i^{(\br)})g'(\mu_i^{(\br)})} &=& 0, \\
    \sum_{i=1}^N \frac{(y_i - \mu_i^{(\br)})\Tilde{Z}_i}{\phi w_i v(\mu_i^{(\br)})g'(\mu_i^{(\br)})} &=& \bD^{-1}\br,
\end{eqnarray}
where $\Tilde{X}_i$ and $\Tilde{Z}_i$ are denote as the $i^{th}$ row of $\Tilde{\bX}$ and $\Tilde{\bZ}$.

Breslow and Clayton [1993] used the Fisher scoring method to solve Equations (A.4) and (A.5) for fixed $\btau$ as an iterated weighted least squares problem involving a working dependent variable and a weight matrix that are updated at each iteration. After determining the maximizers, $\hat{\bgamma}(\btau)$ and $\hat{\br}(\btau)$, for the penalized quasi-likelihood (Equation A.3) under the fixed $\btau$, they employed further approximation to derive the standard REML estimation equation for $\btau$. We briefly summarize the algorithms as follows:

Step 1: Given $\btau$ and $\br$, the fixed effect $\bgamma$ is solved from
\begin{eqnarray}
    (\Tilde{\bX}^T \bV^{-1} \Tilde{\bX})\bgamma = \Tilde{\bX}^T \bV^{-1}\bY^*,
\end{eqnarray}
where $\bY^*$ is a $N \times 1$ working vector having the component $g(\mu_i^{(\br)}) + (y_i - \mu_i^{(\br)})g'(\mu_i^{(\br)})$, $\bV = \bQ^{-1} + \Tilde{\bZ}\bD\Tilde{\bZ}^T$, and $\bQ$ is a $N \times N$ diagonal matrix with diagonal terms $\{\phi w_i v(\mu_i^{(\br)})[g'(\mu_i^{(\br)})]^2\}^{-1}$.

Step 2: For $\hat{\bgamma}$ derived from Equation (A.6), the random effects $\br$ can be estimated by:
\begin{eqnarray}
    \hat{\br} = \bD \Tilde{\bZ}^T \bV^{-1} (\bY^* - \Tilde{\bX}\hat{\bgamma})
\end{eqnarray}

Step 3: Let the component of $\btau$ is $\tau_i$, $\tau_i$ can be obtained from the following estimating equation:
\begin{eqnarray}
    -\frac{1}{2} \left [ (\bY^* - \Tilde{\bX}\hat{\bgamma})^T\bV^{-1}\frac{\partial \bV}{\partial \tau_i} \bV^{-1}(\bY^* - \Tilde{\bX}\hat{\bgamma}) - \mbox{tr} (\bP \frac{\partial \bV}{\partial \tau_i}) \right ] = 0,
\end{eqnarray}
where $\bP = \bV^{-1} - \bV^{-1} \Tilde{\bX} (\Tilde{\bX}^T\bV^{-1}\Tilde{\bX})^{-1} \Tilde{\bX}^T\bV^{-1}$.

Step 4: Update $\bY^*$ at the end of each iteration. The PQL estimators are established once convergence is achieved.

\newpage
\section*{Web Appendix B. Simulation Settings}
\begin{table}[!htbp]
\caption{Parameter setup used in Equation (3.1)} 
\begin{center}
\begin{tabular}{lccc}
\hline
&$k=1$&$k=2$&$k=3$\\
\hline
$\alpha_k^*$ & 0.5 & 1.5 & 1 \\
$\beta_k^*$ & 0.3 & 0.5 & 0.7 \\
$\sigma_k$ & 1 & 1.5 & 2 \\
\hline
\end{tabular}
\end{center}
\end{table}

We calculate the true values of the estimand, the log-RMST for each treatment and subgroup, using Monte Carlo method. By doing this, we fit a RMST regression model on a simulated dataset having single study with a size of 10 million samples, which is sampled from Equation (3.1) with $nt=1$ and $\tau=0$. The RMST regression model is written as follows:
\begin{eqnarray}
log \left( E[T_{i} \wedge t^* | X_{i}, Z_{i}] \right) = \sum_{k=1}^3\alpha_{k}I[Z_{i}=k] + X_{i}\sum_{k=1}^3\beta_kI[Z_{i}=k],
\end{eqnarray}
The true values of the coefficients in Equation (A.9) are shown in Web Table 2.

\begin{table}[!htbp]
\caption{True value of the parameters in Equation (A.9)} 
\begin{center}
\begin{tabular}{lccc}
\hline
&$k=1$&$k=2$&$k=3$\\
\hline
$\alpha_k$ & 0.687 & 1.070 & 0.877 \\
$\beta_k$ & 0.171 & 0.116 & 0.179 \\
\hline
\end{tabular}
\end{center}
\end{table}

\newpage
\begin{figure}[!htbp]
\centering\includegraphics[scale = 0.8]{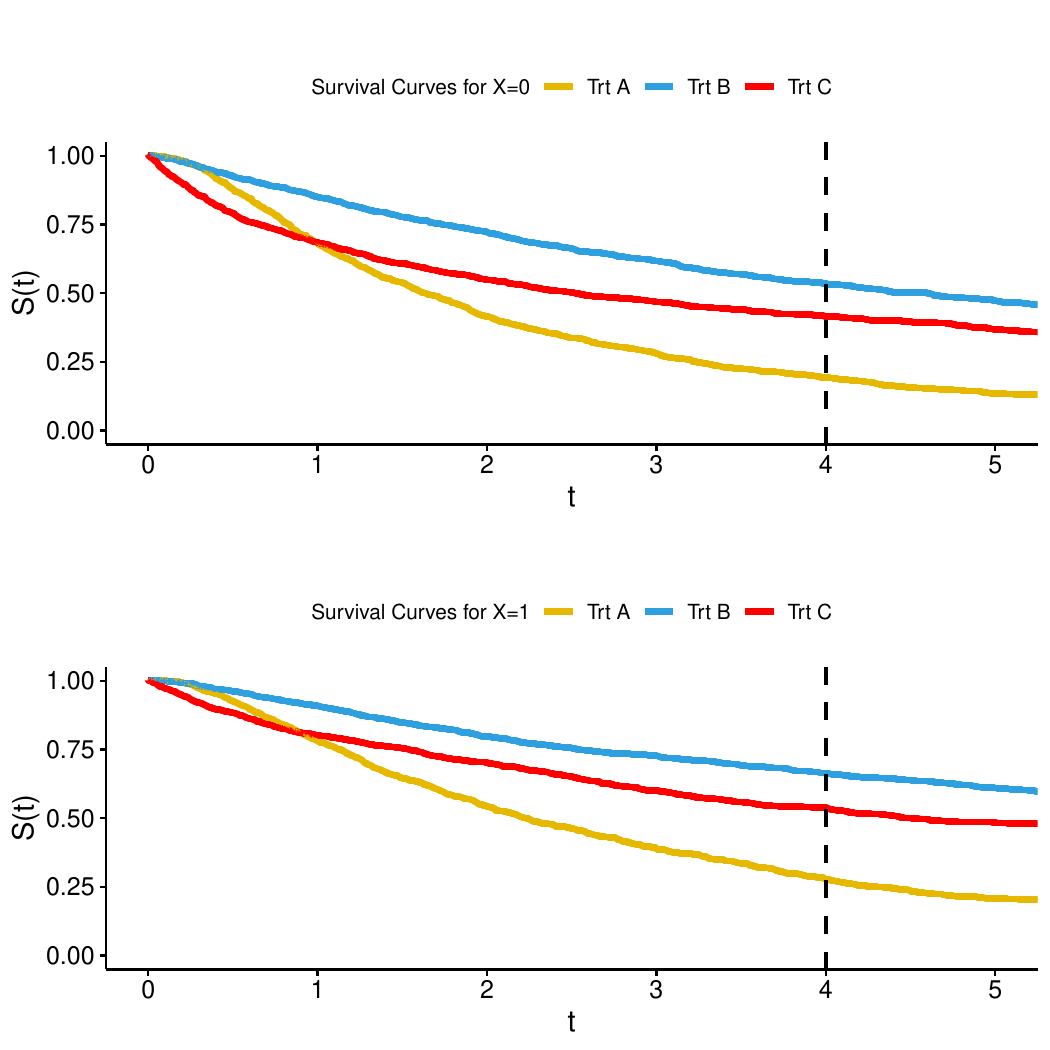}
\caption{Survival curves for three treatments by subgroups $x=0$ and $x=1$, which are considered in the simulation study. These curves are generated from a simulated dataset under Equation (3.1) assuming $nt=1$ and $\tau = 0$ (i.e., fixed effect model) with sample size of 10,000.}
\end{figure}

\newpage
\section*{Web Appendix C. Additional Simulation Study Results}
\begin{figure}[!htbp]
\centering\includegraphics[scale = 0.7]{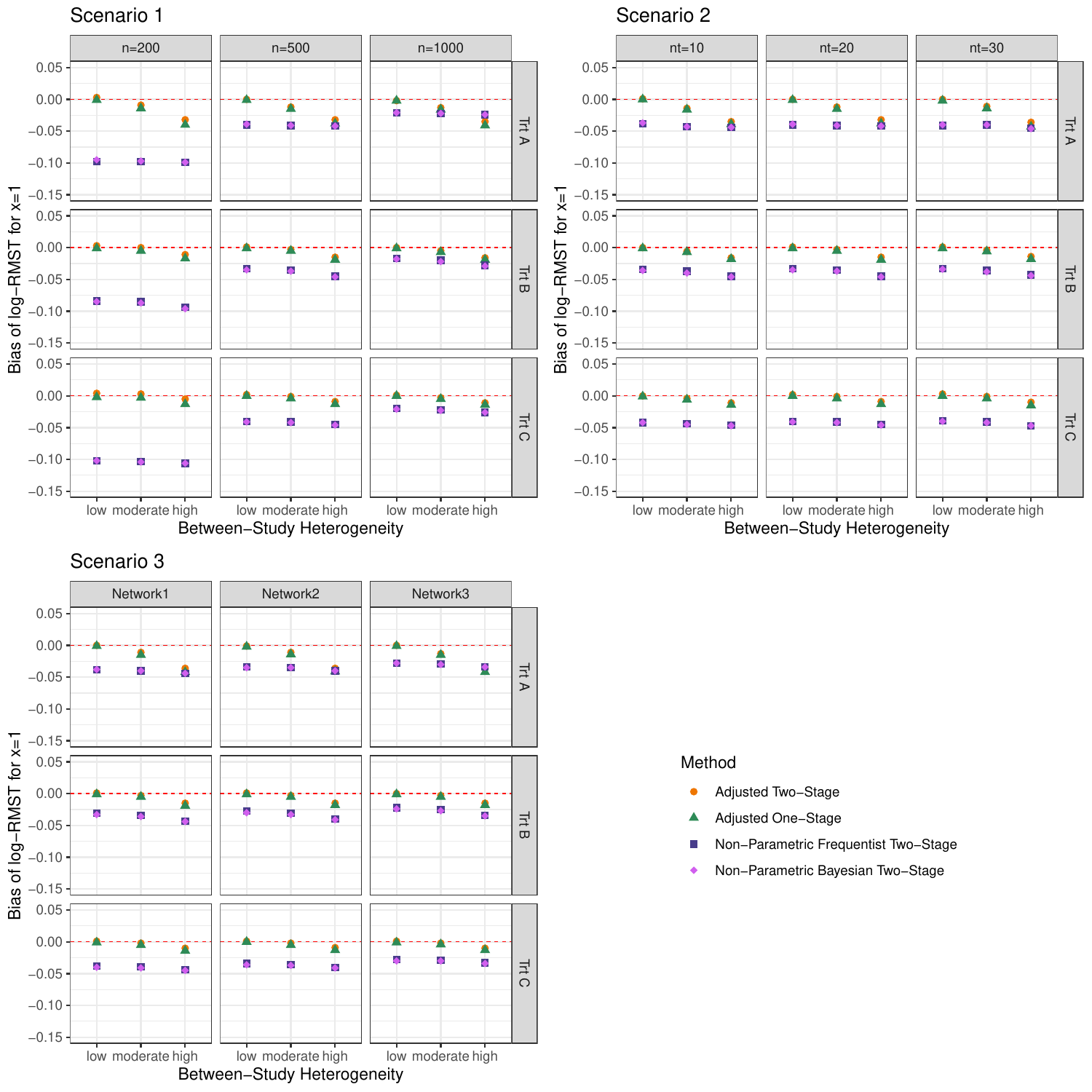}
\caption{Simulation results of bias of log-RMST of three treatments (A, B, and C) for the subgroup with $x=1$ under three scenarios.}
\end{figure}

\newpage
\begin{figure}[!htbp]
\centering\includegraphics[scale = 0.7]{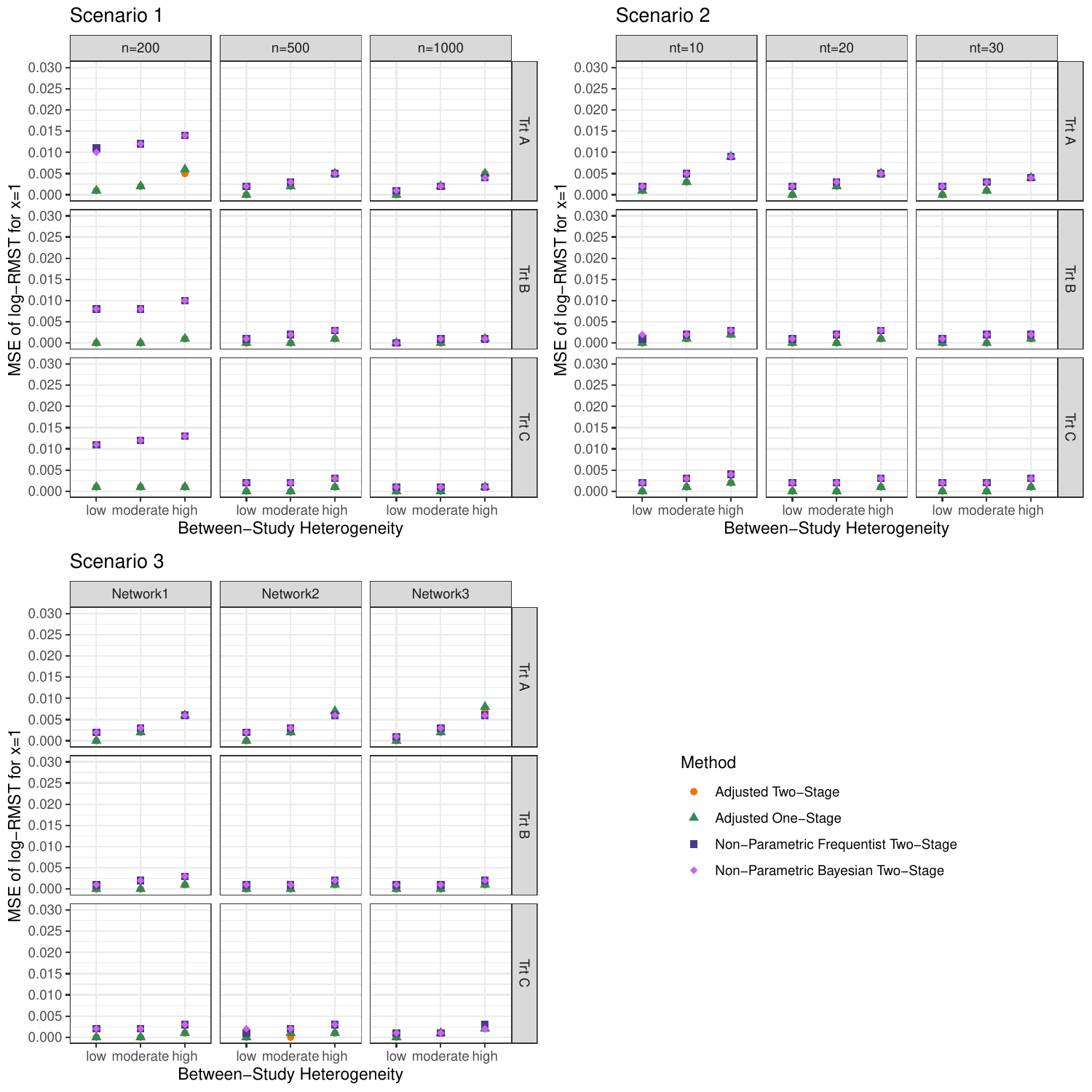}
\caption{Simulation results of mean squared error (MSE) of log-RMST of three treatments (A, B, and C) for the subgroup with $x=1$ under three scenarios.}
\end{figure}

\newpage
\begin{figure}[!htbp]
\centering\includegraphics[scale = 0.7]{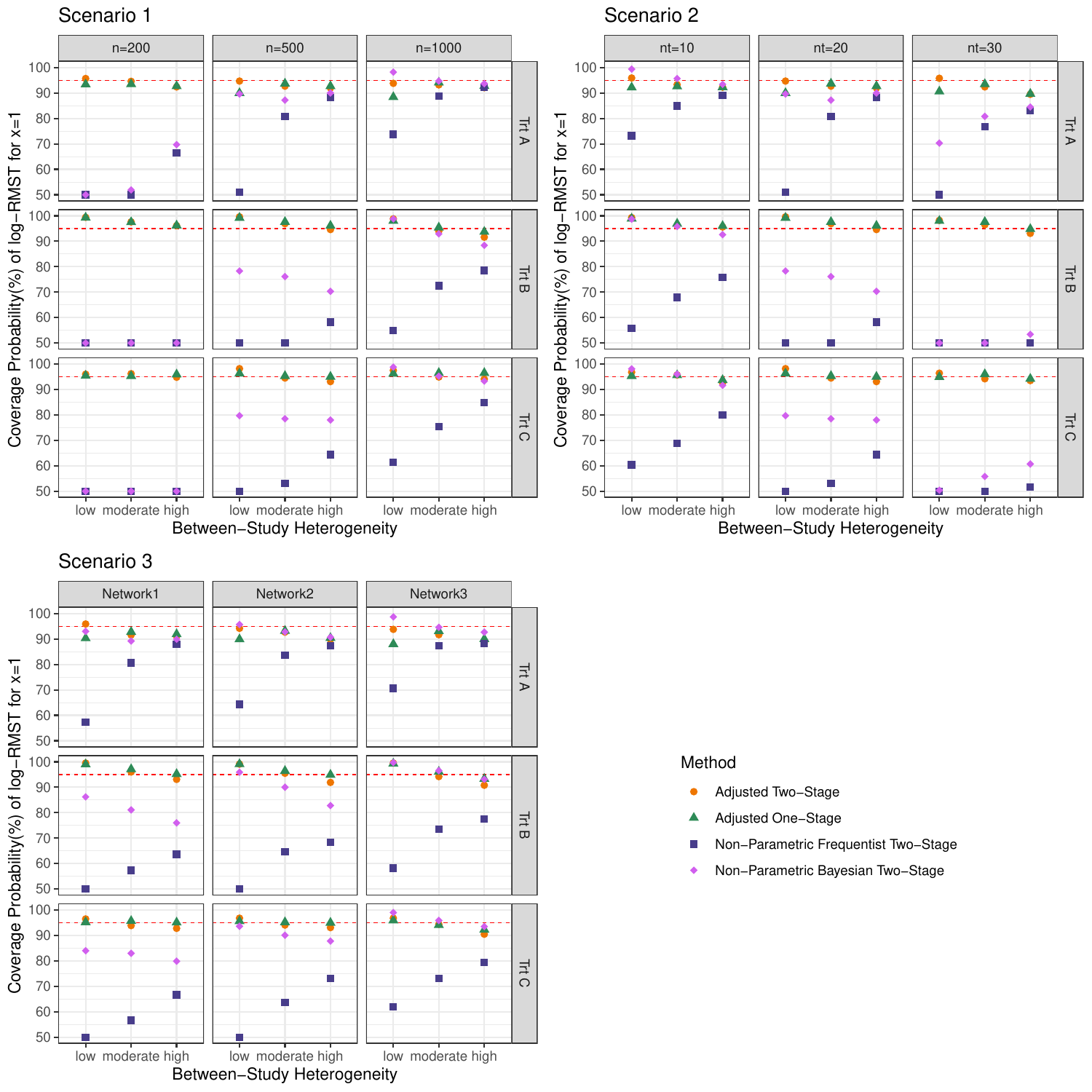}
\caption{Simulation results of coverage probabilities of log-RMST of three treatments (A, B, and C) for the subgroup with $x=1$ under three scenarios. The coverage probabilities below 50\% are truncated to 50\%.}
\end{figure}

\newpage
\section*{Web Appendix D. Additional Data Analysis Results}
\begin{table}[!htbp]
\caption{Data analysis results with the bleeding outcome (top) and the all-cause mortality outcome (bottom) using the proposed two-stage and one-stage RMST NMA-IPD models with a log link function. The coefficient estimates for log RMSTs of treatments (first three rows in each table) and treatment-by-covariate interactions (denoted as treatment:covariate) are presented along with 95\% confidence intervals (CI), Wald test p-values (P-Val), and estimates of standard deviations of random effects ($\tau$) as a measure of between-study heterogeneity.}
\begin{center}
\begin{tabular}{ccccccc}
\hline
\multicolumn{7}{c}{Any Bleeding} \\
\hline
& \multicolumn{3}{c}{Two-Stage} & \multicolumn{3}{c}{One-Stage}\\
Variable & Coef (95\% CI) & P-Val & $\tau$ & Coef (95\% CI) & P-Val & $\tau$\\
\hline
Warfarin & 3.206 (3.145, 3.267) & $<$.01 & 0.061 & 3.207 (3.119, 3.295) & $<$.01 & 0.089\\
SD-DOAC & 3.221 (3.131, 3.311) & $<$.01 & 0.092 & 3.220 (3.118, 3.322) & $<$.01 & 0.104\\
LD-DOAC & 3.266 (3.195, 3.337) & $<$.01 & 0.071 & 3.252 (3.207, 3.297) & $<$.01 & 0.034\\
W:Sex & -0.041 (-0.092, 0.01) & 0.11 & 0.050 & -0.043 (-0.088, 0.002) & 0.06 & 0.043 \\
SD:Sex & -0.003 (-0.064, 0.058) & 0.92 & 0.059 & -0.002 (-0.053, 0.049) & 0.94 & 0.049 \\
LD:Sex & 0.01 (-0.023, 0.043) & 0.56 & 0.026 & 0.006 (-0.025, 0.037) & 0.71 & 0.015 \\
W:Age & -0.007 (-0.009, -0.005) & $<$.01 & 0.002 & -0.007 (-0.009, -0.005) & $<$.01 & 0.001\\
SD:Age & -0.007 (-0.009, -0.005) & $<$.01 & 0.003 & -0.007 (-0.009, -0.005) & $<$.01 & 0.002 \\
LD:Age & -0.006 (-0.008, -0.004) & $<$.01 & 0.002 & -0.007 (-0.009, -0.005) & $<$.01 & 0.001 \\
\hline
\multicolumn{7}{c}{All-Cause Mortality} \\
\hline
& \multicolumn{3}{c}{Two-Stage} & \multicolumn{3}{c}{One-Stage}\\
Variable & Coef (95\% CI) & P-Val & $\tau$ & Coef (95\% CI) & P-Val & $\tau$\\
\hline
Warfarin & 3.409 (3.401, 3.417) & $<$.01 & 0.006 & 3.408 (3.404, 3.412) & $<$.01 & 0.000 \\
SD-DOAC & 3.414 (3.408, 3.42) & $<$.01 & 0.005 & 3.413 (3.407, 3.419) & $<$.01 & 0.003 \\
LD-DOAC & 3.416 (3.41, 3.422) & $<$.01 & 0.003 & 3.417 (3.411, 3.423) & $<$.01 & 0.000 \\
W:Sex & 0.01 (-0.004, 0.024) & 0.15 & 0.011 & 0.011 (0.003, 0.019) & 0.01 & 0.000 \\
SD:Sex & 0.009 (0.001, 0.017) & 0.02 & 0.004 & 0.01 (0.002, 0.018) & 0.01 & 0.000 \\
LD:Sex & 0.013 (-0.001, 0.027) & 0.06 & 0.009 & 0.009 (-0.001, 0.019) & 0.07 & 0.001 \\
W:Age & -0.002 (-0.002, -0.002) & $<$.01 & 0.001 & -0.002 (-0.002, -0.002) & $<$.01 & 0.000 \\
SD:Age & -0.002 (-0.002, -0.002) & $<$.01 & 0.000 & -0.002 (-0.002, -0.002) & $<$.01 & 0.000 \\
LD:Age & -0.002 (-0.002, -0.002) & $<$.01 & 0.000 & -0.002 (-0.002, -0.002)
 & $<$.01 & 0.000 \\
\hline
\end{tabular}
\begin{tablenotes}
\item W: Warfarin.
\item SD: Standard-Dose DOAC.
\item LD: Low-Dose DOAC.
\end{tablenotes}
\end{center}
\end{table}